\documentclass[12pt]{article}

\usepackage{graphicx}
\graphicspath{{figures/}} 
\usepackage{caption}
\usepackage{subcaption}
\usepackage{physics}

\usepackage{amssymb}
\usepackage{amsmath}
\usepackage{dsfont}				
\usepackage{slashed}			
\usepackage[compat=1.0.0]{tikz-feynman}			

\usepackage{jheppub}
\usepackage{orcidlink} 

\newcommand{\mbbone}{\text{\usefont{U}{bbold}{m}{n}1}}
\MakeRobust{\mbbone}

\def\a{\alpha}
\def\b{\beta}
\def\c{\chi}

\def\d{\delta}
\def\D{\Delta}

\def\eps{\varepsilon}
\def\f{\frac}
\def\g{\gamma}

\def\G{\Gamma}
\def\l{\left}
\def\la{\langle}
\def\ra{\rangle}

\def\mc{\mathcal}
\def\m{\mu}
\def\n{\nu}
\def\nn{\nonumber}

\def\p{\partial}
\def\r{\right}
\def\s{\sigma}
\def\t{\theta}

\def\dst{\displaystyle}
\def\be{\begin{equation}}
\def\ee{\end{equation}}
\def\bea{\begin{eqnarray}}
\def\eea{\end{eqnarray}}
\def\ba{\begin{array}}
\def\ea{\end{array}}
\def\bc{\begin{center}}
\def\ec{\end{center}}
\def\bl{\begin{flushleft}}
\def\el{\end{flushleft}}
\def\br{\begin{flushright}}
\def\er{\end{flushright}}
\def\bi{\begin{itemize}}
\def\ei{\end{itemize}}
\def\bt{\begin{tabular}}
\def\et{\end{tabular}}

\def\dst{\displaystyle}
\def\be{\begin{equation}}
\def\ee{\end{equation}}
\def\bea{\begin{eqnarray}}
\def\eea{\end{eqnarray}}
\def\ba{\begin{array}}
\def\ea{\end{array}}
\def\bc{\begin{center}}
\def\ec{\end{center}}
\def\bl{\begin{flushleft}}
\def\el{\end{flushleft}}
\def\br{\begin{flushright}}
\def\er{\end{flushright}}
\def\bi{\begin{itemize}}
\def\ei{\end{itemize}}
\def\bt{\begin{tabular}}
\def\et{\end{tabular}}

\title{Large charge bootstrap with $U(1)$ current probes}

\author{Kasra Kiaee \orcidlink{0009-0001-9383-9491},}
\author{Alexander Monin~\orcidlink{0000-0001-9219-6474}}

\affiliation{University of South Carolina \\
712 Mains St, 404 \\
Columbia SC, 29208

}
\emailAdd{kkiaee@email.sc.edu}
\emailAdd{amonin@mailbox.sc.edu}

\abstract{We study the large-charge bootstrap for conformal field theories with a $U(1)$ symmetry extending the analysis of scalar probes to include conserved currents. We formulate the bootstrap equations and analyze their solutions assuming the existence of a non-trivial macroscopic limit and that the spectrum organizes into a finite number of Regge trajectories.
We show that current probes lead to additional bootstrap constraints that are absent in the purely scalar case, and we classify the resulting solutions.}

\notoc
\begin{document}

\maketitle

\newpage

\tableofcontents

\newpage
\section{Introduction and motivation}

Conformal field theories (CFTs) often exhibit a drastic simplification in sectors of large quantum numbers. Consider a CFT with a conserved $U(1)$ current and a scalar primary $\Phi_Q$ of charge $Q$ with the minimal scaling dimension $\D_Q$ at fixed $Q$. Via the state-operator correspondence, $\Phi_Q(0)$ prepares the ground state on $\mathbb{R}\times \mathbb S^{d-1}$ of charge $Q$, which induces a finite charge density on the cylinder. If this ground state spontaneously breaks the $U(1)$ symmetry, then at large charge its low-energy physics is governed by a universal effective field theory (EFT) of the corresponding Goldstone mode, with a controlled expansion in inverse powers of $Q$~\cite{Hellerman:2015nra,Monin:2016jmo}.
As a consequence, a broad class of observables in the fixed-charge sector can be universally computed in terms of only on a handful of EFT parameters.

%

It is important to stress that the spontaneous symmetry breaking is an assumption about the structure of the charged ground state, and that if the corresponding order parameter is $U(1)$ invariant no description in terms of a superfluid and a Goldstone mode is available. From the CFT point of view, one would like to characterize the large-charge sector directly from conformal symmetry and crossing symmetry of correlation functions, and to understand to what extent the EFT picture is enforced. In other words which alternatives remain compatible with conformal symmetry and crossing without assuming the EFT a priori.

The authors of~\cite{Jafferis:2017zna} addressed this question in $d=3$ for heavy-light four-point functions with scalar light probes, under the assumption that only a finite number $N$ of Regge trajectories contribute in the relevant channel.
In this case crossing and the existence of the macroscopic limit lead to a closed system of relations for the spectrum and fusion coefficients, which can be encoded in terms of finitely many polynomials. In particular, the spectrum $\omega$ of excitations with spin $\ell$ around the ground state is determined by the roots of a characteristic equation with polynomial coefficients
\begin{equation}
x^N+\sum_{k=1}^N (-1)^k x^{N-k} Q_k(z) = 0,
\end{equation}
with
\begin{equation}
x=\omega^2, \quad z = \f{\ell(\ell+1)}{2}.
\end{equation}
For a single trajectory, $N=1$, the solution is unique and coincides with the prediction of the conformal superfluid EFT. While for $N \geq 2$ it was suggested that the space of solutions to the bootstrap equations is broader than that provided by EFT-like models. For two Regge trajectories one of the solutions presented in~\cite{Jafferis:2017zna} naively does not seem to have a quasiparticle interpretation. More generally, any EFT corresponding to the spontaneously broken $U(1)$ symmetry should possess a zero mode, while there is no constraint $Q_N(0)=0$ coming from studying scalar probes. This constraint seems to play an important role in distinguishing crossing-symmetric solutions that can be realized by a local EFT from those that cannot.

This motivates the main goal of the present paper: to extend the large-charge bootstrap of~\cite{Jafferis:2017zna} to vector operators, in particular to the conserved $U(1)$ current. Vector probes introduce new tensor structures and, for conserved currents, Ward identities that constrain the CFT data. Our assumptions are the same as in the scalar case:
\begin{itemize}

\item
A finite number of Regge trajectories.

\item
The existence of a non-trivial macroscopic limit.

\item
The contribution of new primaries and the descendant of the ground state come at the same order.

\end{itemize}
As a result, the bootstrap equations obtained from current probes impose additional conditions on the polynomials that are invisible to scalar correlators. In particular, we find that current probes enforce a non-trivial constraint at $z=0$ (zero spin), namely $Q_N(0)=0$, which matches the EFT expectation associated with the Goldstone shift symmetry. We also show that both two-Regge-trajectory solutions presented in~\cite{Jafferis:2017zna} admit a quasiparticle interpretation and can be obtained from a unitary EFT.

The paper is organized as follows. In Section~\ref{sec:ScalarProbes} we review the large-charge bootstrap with scalar probes and fix conventions. In Section~\ref{sec:VectorProbes} we extend the analysis to vector probes, including conserved currents, and derive the corresponding bootstrap equations. Assuming a finite number of Regge trajectories, in Section~\ref{sec:Solutions} we analyze the solutions of the bootstrap equations and show that conserved-current probes enforce additional constraints. In Section~\ref{sec:EFTRealizations} we consider specific EFT realizations with one Goldstone and one additional light field and show that both two-Regge-trajectory solutions of~\cite{Jafferis:2017zna} can be obtained from these models. We conclude in Section~\ref{sec:Conclusion} and a number of technical derivations are collected in the appendices.

\section{Scalar probes \label{sec:ScalarProbes}}

In this section we review the scalar heavy-light bootstrap setup that underlies our analysis and introduce the conventions that we will use throughout. We begin by recalling the map between flat space and the cylinder and by fixing our normalization of operators and cross-ratios.

\subsection{Conventions}

The map between $d$-dimensional Euclidean space $\mathbb R^d$ and the cylinder $\mathbb R \times \mathbb S^{d-1}$, with coordinates $(\tau, \t^i)$ is realized by\footnote{Everywhere we set the radius of the sphere to be $R=1$.}
\begin{equation}
\label{eq:CylinderMap}
x^a = e^{\tau} n^a (\t), ~~ n^a n^a = 1.
\end{equation}
Given a local operator ${\mc O}^{a_1\dots}_{b_1\dots}(x)$ in Euclidean space with scaling dimension $\D_\mc O$ the corresponding operator on the cylinder is defined as
\begin{equation}
\label{eq:OperatorCylinder}
\hat {\mc O}^{a_1\dots}_{b_1\dots} (\tau, \vec n) = |x|^{\D_{\mc O}+\#a-\#b} {\mc O}^{a_1\dots}_{b_1\dots}(x),
\end{equation}
where $\#a$ and $\#b$ denote the number of upper and lower indices. We also introduce the following notation for a rescaled operator
\begin{equation}
\label{eq:OperatorRescaled}
\tilde {\mc O}^{a_1\dots}_{b_1\dots} (\tau, \vec n) = |x|^{\D_{\mc O}} {\mc O}^{a_1\dots}_{b_1\dots}(x).
\end{equation}
For scalar operators the two definitions coincide.

States on the cylinder are defined via the operator-state correspondence  
\begin{equation}
\label{eq:OperatorState}
| \mc O \ra = \lim_{x\to 0} \mc O(x) | 0 \ra, ~~ 
\end{equation}
with Hermitian conjugation for a scalar primary
\begin{equation}
\label{eq:ScalarHermitian}
\la \phi | = \lim_{x\to \infty} x^{2\D} \la 0 | \phi(x),
\end{equation}
and similarly for a spin $\ell$ operator (a rank-$\ell$ traceless symmetric tensor)
\begin{equation}
\la T^{a_1\dots a_\ell} | = \lim_{x\to \infty} x^{2\D} I^{a_1 b_1} \dots I^{a_\ell b_\ell} \la 0 | T^{b_1\dots b_\ell}(x),
\end{equation}
with
\begin{equation}
I^{a b} = \d^{ab}-2 n^a n^b.
\end{equation}

We define the conformal cross-ratios
\begin{equation}
u = \f{x_{12}^2x_{34}^2}{x_{13}^2x_{24}^2}, ~~ 
v = \f{x_{14}^2 x_{23}^2}{x_{13}^2x_{24}^2}.
\end{equation}
We will also trade $(u,v)$ for the cylinder coordinates $(\tau,\t)$, for which the angle $\t$ can be written as
\begin{align}
\label{eq:CylinderCoordinates}
\tau & = \f{1}{2}\log u, \\
\t &= \arccos \f{1+u-v}{2\sqrt{u}}.
\end{align}
We also use the following conventions for the standard conformal structures
\begin{align}
V^a_{1,23} = \f{x_{12}^2x_{13}^2}{x_{23}^2} \l ( \f{x_{21}^a}{x_{12}^2} - \f{x_{31}^a}{x_{13}^2} \r ), ~~ 
H^{ab}_{ij}  = g^{ab} x_{ij}^2 - 2x_{ij}^ax_{ij}^b.
\end{align}

\subsection{Four-point functions and crossing}

The starting point of the large-charge bootstrap procedure is the crossing equation. Any four-point function of scalar operators is determined by a single unknown function, and it can be represented in several equivalent forms, each suitable for an expansion in a given channel. We will only need the $s$- and $u$-channel representations, which are related by simultaneously interchanging coordinates $x_1 \leftrightarrow x_4$ and the scaling dimensions $\D_1 \leftrightarrow \D_4$, leading to
\begin{align}
\label{eq:4ptFunctionSSSS-generic}
\la \phi_4(x_4) \phi_3 (x_3) \phi_2 (x_2) \phi_1(x_1) \ra 
& = \f{G^{(s)}(u,v)}{x_{12} ^{\D_1+\D_2} x_{34} ^{\D_3+\D_4}}
\l( \f{x_{24}}{x_{14}}\r )^{\D_1-\D_2}\l( \f{x_{14}}{x_{13}}\r )^{\D_3-\D_4} \\
& = \f{G^{(u)}\l( \f{1}{u},\f{v}{u}\r )}{x_{24} ^{\D_2+\D_4} x_{13} ^{\D_1+\D_3}}
\l( \f{x_{12}}{x_{14}}\r )^{\D_4-\D_2}\l( \f{x_{14}}{x_{34}}\r )^{\D_3-\D_1}, \nn
\end{align}
where $u,v$ are the cross-ratios above.

The functions $G^{(s)}$ and $G^{(u)}$ are one and the same function with interchanged $\D_1 \leftrightarrow \D_4$, namely,
\begin{equation}
G^{(s)} = G^{(4,3,2,1)}, ~~ G^{(u)} = G^{(1,3,2,4)}.
\end{equation}
One immediate consequence of \eqref{eq:4ptFunctionSSSS-generic} is the crossing equation
\begin{equation}
G^{(u)}\l ( \f{1}{u}, \f{v}{u} \r ) = G^{(s)}(u,v) u^{-\f{\D_1+\D_4}{2}}.
\end{equation}
It is convenient to introduce
\begin{equation}
\label{eq:gScalar}
g^{(s)}(u,v) = G^{(s)}(u,v) u^{-\f{\D_1+\D_4}{4}},
\end{equation}
which implies 
\begin{equation}
g^{(u)}(u,v) = G^{(u)}(u,v) u^{-\f{\D_1+\D_4}{4}},
\end{equation}
and the crossing equation becomes\footnote{Equivalently, this is the constraint on $g^{(i,j,k,\ell)}$
\begin{equation}
g^{(1,3,2,4)}\l ( \f{1}{u}, \f{v}{u} \r ) = g^{(4,3,2,1)}(u,v).
\end{equation}}
\begin{equation}
\label{eq:CrossingScalar}
g^{(u)}\l ( \f{1}{u}, \f{v}{u} \r ) = g^{(s)}(u,v).
\end{equation}
Solving this equation in general is an insurmountable task. Help comes from conformal symmetry, in particular from the operator-state correspondence.

\subsection{Cylinder representation and heavy-light expansion}

Considering the limit $x_1\to 0$ and $x_4\to \infty$ and using
\begin{equation}
\tilde {\mc O} (\tau,\vec n) = e^{H \tau} \tilde {\mc O} (\vec n) e^{-H \tau},
\end{equation}
together with \eqref{eq:OperatorRescaled}, \eqref{eq:OperatorState}, and \eqref{eq:ScalarHermitian} we derive
\begin{align}
G^{(s)}(u,v) = \la \phi_4| \tilde \phi_3 (\vec n_3) e^{-H(\tau_3-\tau_2)} \tilde \phi (\vec n_2) | \phi_1 \ra.
\end{align}
Using $u = z \bar z = e^{2\tau}$, we can rewrite \eqref{eq:gScalar} in terms of $z,\bar z$ and expand in a complete set of states, obtaining
\begin{align}
\label{eq:4ptScalarS-channelOPE}
g^{(s)}(z,\bar z) = \sum_E u^{\f{E}{2}-\f{\D_1+\D_4}{4}} \la \phi_4 | \tilde \phi_3 (\vec n_3) | E \ra 
\la E | \tilde \phi_2 (\vec n_2) | \phi_1 \ra, ~~ |z|<1.
\end{align}

Further simplification is only possible once additional assumptions are made about the spectrum of states appearing in the sum. For the large-charge bootstrap the states $| \phi_1 \ra$ and $| \phi_4 \ra$ correspond to the large-charge $Q$ operator of smallest scaling dimension $\D_Q$. The probe operators $\phi_2$ and $\phi_3$ are considered "light", meaning their scaling 
dimensions are much smaller than $\D_Q$. In~\cite{Jafferis:2017zna} charged probes were considered. We instead review the procedure for neutral operators, since later we will need precisely those when we start adding non-scalar probes. As a result we are interested in\footnote{Here we traded the superscript $(s)$ for $Q$. Clearly, the $u$-channel function is given by $g^{-Q}$.}
\begin{align}
\label{eq:gScalarS-channelOPE}
g^{Q}(z,\bar z) = \sum_E u^{\f{E-\D_Q}{2}} \la Q | \tilde \phi_3 (\vec n_3) | E \ra 
\la E | \tilde \phi_2 (\vec n_2) | Q \ra,
\end{align}
with neutral operators $\phi_2$ and $\phi_3$ and the sum running over states with charge $Q$.

The main observation of the large-charge bootstrap is that the terms in the sum \eqref{eq:gScalarS-channelOPE} can be organized in a power series in the charge $Q$. It is assumed that at leading order only the state $| Q \ra$ contributes to the sum. The relative $Q$-suppression of the first descendant of $|Q\ra$ is fixed by conformal symmetry, and in the case of neutral operators, is given by $\D_Q^{-1}$ (see Appendix~\ref{app:VectorDetails}). It is assumed that other primaries, with scaling dimensions
\begin{equation}
E=\D_Q+\omega_\ell,
\end{equation}
whose descendants are even further suppressed, contribute at this order as well. As a result, the ansatz for the four-point function is
\begin{equation}
\label{eq:g-Scalar}
g^Q (\tau,\t) = | \lambda_{S,\D_Q}|^2 \l [1 + \f{\d^2}{2 \D_Q} h(\tau,\t) \r ],
\end{equation}
with
\begin{align}
\label{eq:f-Anzatz-Scalars}
h(\tau, \t) & = e^{\tau} \cos\t 
+ \sum_{\mc O_{E,\ell}} |\m_{S,\mc O}|^2 \, e^{\omega_\ell \tau} \, C^{(d/2-1)}_\ell(\cos\t), ~~ \cos\t = \vec n_2 \vec n_3, ~~ \tau <0,
\end{align}
where $C_\ell^{(d/2-1)}(\cos\t)$ are Gegenbauer polynomials, and we have removed the superscript $Q$, since all dependence on $Q$ only enters trough $\D_Q$. Since the crossing equation \eqref{eq:CrossingScalar} is trivially satisfied at leading order, the constraint on the subleading term becomes
\begin{equation}
\label{eq:CrossingScalars}
h(-\tau,\t) = h(\tau, \t).
\end{equation}

For $\ell \geq 2$ the number of operators (Regge trajectories) for each spin is the same $N_\ell=N$, while this number can be different for $\ell=0,1$, however, it cannot exceed $N$ (see the comment after Eq.~\eqref{eq:GeneratingScalar}). As a result, we can write
\begin{align}
\label{eq:f-Anzatz-Scalars-NRegge}
h(\tau, \t) & = \sum_{i=1}^N \sum_{\ell} |\m_{S,i,\ell}|^2 \, e^{\omega_{i,\ell} \tau} \, C^{(d/2-1)}_\ell(\cos\t),
\end{align}
with 
\begin{equation}
\omega_{i,\ell} = E_{i,\ell}-\D_Q.
\end{equation}
The descendant contribution
\begin{equation}
e^{\tau} \cos\t = \f{e^{\tau}}{d-2} C_1^{(d/2-1)}(\cos\t),
\end{equation}
is included in the sum. For a non-degenerate case, when only for one Regge trajectory, say number 1,
\begin{equation}
\label{eq:ScalarNonDegenerateOmega}
\omega_{1,1} = 1,
\end{equation}
it implies that
\begin{equation}
\label{eq:ScalarNonDegenerateMu}
|\m_{S,1,1}|^2 = \f{1}{d-2}.
\end{equation}
For a degenerate case the descendant contribution corresponds to a combination of coefficients.

\subsection{Macroscopic limit and singular behavior}

The function $g^Q(\tau,\t)$ has a singularity at $\tau=\t=0$ controlled by the $t$-channel expansion. Since we do not have access to the light-light OPE, we do not know the exact behavior of the function close to this point. However, in the large charge limit certain terms in the $t$-channel expansion are $Q$-enhanced compared to others, and therefore, due to the order of limits, it is precisely these operators that control the small-distance expansion around the singularity in $h(\tau,\t)$.

From the EFT perspective, this means that at distances much smaller than the IR scale (the radius of the sphere), but larger than the UV scale defined by the charge density (chemical potential $\m^{d-1} \sim Q$), the superfluid description can still be used to derive the asymptotic behavior of the four-point function. Going to yet smaller distances we start resolving the "discrete" nature of the superfluid. In other words, at truly short distances the asymptotic behavior is controlled, as it should, by the lightest operator in the $t$-channel.

The way to determine the asymptotic behavior in the large-charge limit is to demand the existence of a macroscopic limit. Due to conformal symmetry this limit is equivalent to
\begin{equation}
Q\to\infty, ~~ \tau,\t \to 0, ~~ \m \tau, \, \m \t = \text{fixed}.
\end{equation}
The existence of the macroscopic limit and \eqref{eq:g-Scalar} imply that the leading singularity of the four-point function at 
$\tau=\t=0$ is bounded by\footnote{In~\cite{Jafferis:2017zna} the most singular term was written as
$(\tau^2+\t^2)^{-1}$, which is equivalent to our representation
\begin{equation}
\f{1}{\tau^2+\t^2} = \f{1}{\tau^2} \l ( \f{\tau^2}{\tau^2+\t^2} \r ) = \f{\text{Regular}}{\tau^2}.
\end{equation}}
\begin{equation}
\label{eq:ShortDistansScalar}
h(\tau,\t) \sim \f{1}{\tau^d}.
\end{equation}
Even though, contrary to the charged case considered in~\cite{Jafferis:2017zna}, there are non-trivial solutions for the crossing (continuity) equation for a less singular behavior, $h(\tau,\t) \sim 1/\tau^{d-2}$, the macroscopic limit trivializes, and in this paper we only consider solutions corresponding to a non-trivial macroscopic limit and thus saturating the bound. As a result, the four-point function can be approximated by
\begin{equation}
\label{eq:AsymptoticScalars}
h(\tau,\t) = \f{1}{\tau^{d}}B_d \l( \f{\tau}{\sqrt{\tau^2+\t^2}}\r) + \f{1}{\tau^{d-1}}B_{d-1} \l( \f{\tau}{\sqrt{\tau^2+\t^2}}\r)+\text{less singular}.
\end{equation}

\subsection{Integral constraints and recurrence relation}

The analytic structure of the four-point function thus suggests the following way to solve the crossing equation. We consider the following integrals (all even derivatives vanish automatically due to the parity of $h(\tau, \t)$)
\begin{equation}
I_{n} = \lim_{\eps\to 0}\int _0^\pi\l [\p_\tau^{2n-1} h (-\eps,\t)-\p_\tau^{2n-1} h (\eps,\t) \r ] C_\ell^{(d/2-1)}(\cos \t) \sin^{d-2} \t \, d\t, \quad n=1, 2, \dots
\end{equation}
On one hand these integrals can be evaluated using the ansatz \eqref{eq:f-Anzatz-Scalars} for $\tau<0$ and the crossing equation \eqref{eq:CrossingScalars} leading to 
\begin{equation}
I_{n} = 2 \sum_{i=1}^{N}|\m_{S,i,\ell}|^2 \omega_{i,\ell}^{2n-1} n_{d,\ell},
\end{equation}
with the norm of the Gegenbauer polynomials given by
\begin{equation}
n_{d,\ell} = \l [ C_\ell^{(d/2-1)}(\cos \t) \r ]^2 \sin^{d-2} \t \, d\t = \f{2^{4-d} \pi \G(\ell+d-2)}{(2\ell+d-2)\ell! \, \G\l( \f{d}{2} - 1\r )}.
\end{equation}
On the other hand, due to the smoothness of the four-point function away from the singularity, it is clear that the integral is saturated around $\tau=\t=0$, and we can use the asymptotic behavior \eqref{eq:AsymptoticScalars} to evaluate it, leading to
\begin{equation}
I_{n} = \lim_{\eps \to 0}\sum_{m=0}^\infty\f{\eps^{d-2}\eps^{2m}\eps}{\eps^{d-2k}\eps^{2n-1}} C_\ell^{(d/2-1)}(1) 
\tilde P_{m}(J^2_{d,\ell}) = \f{(d-2)_\ell}{\ell!} \tilde P_{n}(J^2_{d,\ell}),
\end{equation}
where used the Pochhammer 
\begin{equation}
(a)_n = \f{\G(a+n)}{\G(a)},
\end{equation}
and the fact that around $\t=0$, the Gegenbauer polynomials admit an expansion in powers of $\t$ with coefficients that are themselves polynomials of $J^2_{d,\ell}$
\begin{equation}
C_\ell^{(d/2-1)}(\cos \t) = \f{(d-2)_\ell}{\ell!} \l [1 + \sum_m \a_m (J^2_{d,\ell} ) \t^{2m} \r ],
\end{equation}
and
\begin{equation}
J^2_{d,\ell} = \f{\ell(\ell+d-2)}{d-1}.
\end{equation}
As a result, we have to solve the following system of equations
\begin{equation}
(d-2)\sum_{i=1}^{N}|\m_{S,i,\ell}|^2 \omega_{i,\ell}^{2n-1} = \f{2\ell+d-2}{d}P_{n}(J^2_{d,\ell}).
\end{equation}
Introducing the notation
\begin{align}
z& = J^2_{d,\ell}, ~~ x_i (z) = \omega_{i,\ell}^2, ~~ |\m_{S,i,\ell}|^2 = \f{2\ell+d-2}{d(d-2)} A_i (z) \omega_{i,\ell},
\end{align}
we can rewrite the equations as
\begin{equation}
\label{eq:RecurrenceSolution}
\sum_{i=1}^{N} A_i(z) x_i^n(z)  = P_{n}(z), ~~ n=1, 2, \dots,
\end{equation}
with at least one root
\begin{equation}
x_1(1) = 1.
\end{equation}
For $z >1$ (equivalently, for spin $\ell>1$) this system of equations can be viewed as a solution to the following  order-$N$ recurrence
\begin{equation}
\label{eq:RecurrenceScalar}
P_{n+N}(z)= \l[ x_1(z)+\dots+x_{N}(z) \r ] P_{n+N-1}(z) + (-1)^N \dots + x_1(z) \dots x_{N}(z) P_{n}(z).
\end{equation}
Indeed, introducing the shift operator $E$ acting as
\begin{equation}
E P_{n}(z) = P_{n+1}(z),
\end{equation}
we can rewrite the recurrence as
\begin{equation}
(E-x_1) \dots (E-x_{N}) P_n = 0,
\end{equation}
immediately showing that \eqref{eq:RecurrenceSolution} is a solution. 

Consistency of \eqref{eq:RecurrenceScalar} demands that $x_i(z)$ be roots of a characteristic equation with polynomial coefficients
\begin{equation}
x^{N} - Q_1(z) x^{N-1} + \dots + Q_{N}(z) x + Q_{N}(z) = 0.
\end{equation}
Specifying the polynomials $Q_{i}(z)$ we can find the spectrum and providing the first $N$ polynomials $P_{i}(z)$ we can find the rest of $P_n(z)$. Therefore, any solution is characterized by two sets of polynomials: $Q_{i}(z)$ and $P_{i}(z)$, with $i=1,\dots, N$.

We can make one more observation about the number of $\ell=0, 1$ operators. For $z>1$ the generating function of the recurrence is 
\begin{equation}
\label{eq:GeneratingScalar}
W(t,z) = \sum_{n=1}^\infty P_n(z) t^n = \sum_{i=1}^{N} \f{A_i(z) t}{1-t x_i(z)}.
\end{equation}
Its analytic structure demands that there be at most $N$ poles in $t$ for any fixed $z$. Therefore, the number of states with spin $\ell=0,1$ cannot exceed~$N$.

\section{Vector probes \label{sec:VectorProbes}}

In this section we extend the large-charge bootstrap setup of the previous section to correlators with vector probes, and derive the corresponding coupled system of crossing/continuity equations. 

Concretely, we write the most general conformally covariant four-point functions with one and with two vector probes, in direct analogy with \eqref{eq:4ptFunctionSSSS-generic}. In contrast to the scalar case, which is controlled by a single function of the cross ratios, the mixed scalar-vector correlator involves two independent tensor structures functions, while the vector-vector correlator involves five of those.

Next, we perform the $s$-channel conformal-block expansion on the cylinder keeping terms up to order $\D_Q^{-1}$. This produces the analogues of \eqref{eq:f-Anzatz-Scalars}: each structure decomposes into the leading ground-state contribution and the first subleading contributions from descendants and additional primaries, with a set of a priori unknown coefficients.

Finally, imposing the continuity equations on the resulting set of functions (together with the scalar sector) yields the desired bootstrap system. Different linear combinations of tensor structures are equivalent. However, we choose a basis that is particularly convenient because crossing symmetry acts diagonally on the corresponding functions, so that different structures do not mix.

The relevant tensor structures were derived in~\cite{Costa:2011mg,Costa:2011dw} and are now standard components of the CFT dictionary. We therefore restrict the discussion here to what is needed for the bootstrap analysis and defer the derivations to Appendix~\ref{app:VectorDetails}.

\subsection{Four-point functions}

The four-point function involving one scalar and one vector probe with scaling dimensions $\D_S$ and $\D_V$ correspondingly can be written as
\begin{align}
\la \Phi_{-Q}(x_4) \phi (x_3) V^{a} (x_2) \Phi_Q (x_1) \ra 
& = \f{F^Q_4 (u,v) V_{2,31}^a + F^Q_1 (u,v) V_{2,34}^a}{x_{12} ^{\D_Q+\D_V+1} x_{34} ^{\D_S+\D_Q}}
\l( \f{x_{24}}{x_{14}}\r )^{\D_Q-\D_V-1}\l( \f{x_{14}}{x_{13}}\r )^{\D_S-\D_Q},
\end{align}
and for two vector probes we have
\begin{align}
\label{eq:4ptFunctionSVVS-Q}
\la \Phi_{-Q}(x_4) V^a (x_3) V^{b} (x_2) \Phi_Q(x_1) \ra & = \f{1}{x_{12} ^{\D_Q+\D_V+1} x_{34} ^{\D_V+\D_Q+1}}
\l( \f{x_{24}}{x_{14}}\r )^{\D_Q-\D_V-1}\l( \f{x_{14}}{x_{13}}\r )^{\D_V-\D_Q+1} \nn \\
& \qquad \Big [ x_{23}^2 H_{23}^{ab} F^Q_{23} + F^Q_{11} (u,v) V^a_{3,24}V_{2,34}^b 
+ F^Q_{14} (u,v) V^a_{3,24}V_{2,31}^b \nn \\
& \qquad \qquad + F^Q_{41} (u,v) V^a_{3,21}V_{2,34}^b + F^Q_{44} (u,v) V^a_{3,21}V_{2,31}^b \Big ].
\end{align}
The most convenient basis for implementing the bootstrap procedure is the following. For scalar-vector
\begin{align}
u^{\f{\D_Q}{2}} H^Q_2 & = \f{1-u-v}{2u} F^Q_1 + \f{1}{2} (1-u-v) F^Q_4, \\
u^{\f{\D_Q}{2}} H^Q_3 & = \f{1}{\sqrt{u}} F^Q_1 + \sqrt{u} F^Q_4
\end{align}
and for two vectors
\begin{align}
u^{\f{\D_Q}{2}} H^Q_{\d} &= \f{v}{\sqrt{u}} F^Q_{23}, \\
u^{\f{\D_Q}{2}} H^Q_{23} & = F^Q_{11}+u F^Q_{14}+\f{1}{u}F^Q_{41}+F^Q_{44}+2 F^Q_{23}, \nn\\
u^{\f{\D_Q}{2}} H^Q_{22}& = \f{1-u-v}{2\sqrt{u}} F^Q_{11} + \f{1}{2} \sqrt{u} (1-u+v) F^Q_{14} + 
\f{1-u-v}{2u^{3/2}} F^Q_{41} + \f{1-u+v}{2\sqrt{u}} F^Q_{44}+\f{1-u}{\sqrt{u}}F^Q_{23}, \nn \\
-u^{\f{\D_Q}{2}}  H^Q_{33}& = \f{1-u+v}{2\sqrt{u}} F^Q_{11} + \f{1}{2} \sqrt{u} (1-u+v) F^Q_{14} + 
\f{1-u-v}{2u^{3/2}} F^Q_{41} + \f{1-u-v}{2\sqrt{u}} F^Q_{44}+\f{1-u}{\sqrt{u}}F^Q_{23},  \nn \\
-u^{\f{\D_Q}{2}} H^Q_{32}& = 
\f{1+u^2-(1+u)v}{2 u} F^Q_{11} + \f{1}{2} \Big[ 1+(u-v)^2 \Big ]F^Q_{14} \nn \\
& \qquad 
+ \f{u^2+(1-v)^2}{2u^{2}} F^Q_{41} + \f{1+u^2-(1+u)v}{2 u} F^Q_{44} 
+ \f{2(1+u^2)-3(1+u)v+v^2}{2 u}F^Q_{23}.
\nn
\end{align}
With this choice, crossing symmetry acts simply on the $H$-functions
\begin{equation}
\label{eq:CrossingH}
H^{-Q}_{\{3,23,\d,32\}}\l ( \f{1}{u}, \f{v}{u} \r ) = H^{Q}_{\{3,23,\d,32\}}(u,v), \qquad
H^{-Q}_{\{2,22,33\}}\l ( \f{1}{u}, \f{v}{u} \r ) = - H^{Q}_{\{2,22,33\}}(u,v).
\end{equation}

As in the scalar case we assume that at leading order in $Q$ only the operator corresponding to the ground state $| Q \ra$ contribute to the heavy-light OPE. We denote the corresponding fusion coefficients as $\lambda_{S,Q}$ and $\lambda_{V,Q}$.
The descendant and new primaries appear at next to leading order with the relative suppression $\D_Q^{-1}$. Using the cylinder coordinates \eqref{eq:CylinderCoordinates} we obtain for scalar-vector
\begin{align}
H^Q_2(\tau,\t) & = \lambda_{V,Q}^* \lambda_{S,Q} 
\l [ 1 + \f{\d_S \d_V}{2 \D_Q} h_{2} (\tau,\t) \r ] \\
H^Q_3(\tau,\t) & = \lambda_{V,Q}^* \lambda_{S,Q} \, \f{\d_S \d_V}{2 \D_Q} h_{3} (\tau,\t), \nn
\end{align}
and for two vectors
\begin{align}
H^Q_{23} & = \f{|\lambda_{V,Q}|^2 \d_V^2}{2\D_Q} h_{23}(\tau,\t), \qquad H^Q_{\d} = \f{|\lambda_{V,Q}|^2 \d_V^2}{2\D_Q} h_{\d}(\tau,\t), \nn \\
H^Q_{22} & = \f{|\lambda_{V,Q}|^2 \d_V^2}{2\D_Q} h_{22}(\tau,\t), \qquad H^Q_{33} = \f{|\lambda_{V,Q}|^2 \d_V^2}{2\D_Q} h_{33}(\tau,\t), \nn \\
H^Q_{33} & = |\lambda_{V,Q}|^2 \l [ 1 + \f{ \d_V^2}{2\D_Q} h_{32}(\tau,\t)\r ].
\end{align}
At leading order in $Q^{-1}$ the crossing equations \eqref{eq:CrossingH} impose the following constraints
\begin{align}
\label{eq:Conserved}
|\lambda_{S,-Q}|^2 = |\lambda_{S,Q}|^2, \quad |\lambda_{V,-Q}|^2 = |\lambda_{V,Q}|^2, \quad \lambda_{V,-Q}^*\lambda_{S,-Q} = -\lambda_{V,Q}^*\lambda_{S,Q}.
\end{align}
From this point forward we restrict to the conserved current corresponding to the $U(1)$ symmetry
as the probe $V^a$. In this case the leading order conditions \eqref{eq:Conserved} are obviously satisfied, since in this case 
$\lambda _{V,Q} \sim Q$.

Assuming that fusion coefficients corresponding to the subleading primaries have the same $Q$-dependence as the leading order operator we find the following crossing equations for $h$-functions\footnote{We omitted the superscript $Q$ since all $Q$-dependence was factored out in $\lambda_{V,Q}$. Also note the opposite parity of $h_{2}$ and $h_{3}$ as compared to $H_{2}$ and $H_{3}$ in \eqref{eq:CrossingH}.}
\begin{equation}
\label{eq:Crossingh}
h_{\{2,23,\d,32\}} (-\tau, \t) = h_{\{2,23,\d,32\}}(\tau,\t), \qquad
h_{\{3,22,33\}} (-\tau, \t) = - h_{\{3,22,33\}}(\tau,\t).
\end{equation}
The $s$-channel expansion ($\tau<0$) of these functions is given by (see Appendix~\ref{app:VectorDetails})
\begin{align}
\label{eq:hSVConserved}
h_{2} (\tau, \t) & = e^{\tau} \cos \t  + \sum_{\ell=1}^\infty \m_{S,\ell} \, \n^+_{\ell} \, \f{\ell+d-2}{\ell+d-2-\omega_\ell} 
e^{\omega_\ell \tau} \, C_{\ell}^{(d/2-1)}(\cos \t), \\
h_{3} (\tau, \t) & = \f{e^{\tau}}{d-1} + (d-2) \sum_{\ell=1}^\infty \m_{S,\ell} \, \n^+_{\ell} \, \f{\omega_\ell}{\ell(\ell+d-2-\omega_\ell)} e^{\omega_\ell \tau} \, C_{\ell-1}^{(d/2)}(\cos \t),
\end{align}
and for vectors
\begin{align}
\label{eq:hVVConserved}
h_{23} (\tau, \t) & = d(d-2) \sum_{\ell=2}^\infty
\l | \n^+_{\ell} \r |^2 \f{\omega_\ell^2}{\ell^2(\ell+d-2-\omega_\ell)^2} \,
e^{\omega_\ell \tau} \, C_{\ell-2}^{(d/2+1)}(\cos \t), \\
h_{\d} (\tau, \t) & = \f{e^{\tau}}{(d-1)^2} + (d-2) \sum_{\ell=1}^\infty
\l | \n^+_{\ell} \r |^2 \f{\omega_\ell^2}{\ell^2(\ell+d-2-\omega_\ell)^2} \,
e^{\omega_\ell \tau} \, C_{\ell-1}^{(d/2)}(\cos \t), \nn \\
h_{22} (\tau, \t) & = \f{e^{\tau}}{d-1} + (d-2) \sum_{\ell=1}^\infty
\l | \n^+_{\ell} \r |^2 \f{\omega_\ell (\ell+d-2)}{\ell(\ell+d-2-\omega_\ell)^2} \,
e^{\omega_\ell \tau} \, C_{\ell-1}^{(d/2)}(\cos \t), \nn \\
h_{33} (\tau, \t) & =  \f{e^{\tau}}{d-1} + (d-2) \sum_{\ell=1}^\infty
\l | \n^+_{\ell} \r |^2 \f{\omega_\ell (\ell+d-2)}{\ell(\ell+d-2-\omega_\ell)^2} \,
e^{\omega_\ell \tau} \, C_{\ell-1}^{(d/2)}(\cos \t), \nn \\
h_{32} (\tau, \t) & = e^{\tau} \cos \t + \sum_{\ell=1}^\infty
\l | \n^+_{\ell} \r |^2 \f{(\ell+d-2)^2}{(\ell+d-2-\omega_\ell)^2} \,
e^{\omega_\ell \tau} \, C_{\ell}^{(d/2-1)}(\cos \t). \nn
\end{align}
The reason equations corresponding to the correlators with the current only involve operators with non-zero spin is the Ward identity, implying that only three point functions with two identical scalars and the conserved current are non-zero.

\section{Solutions \label{sec:Solutions}}

In the scalar case, solving the continuity equations required the asymptotic behavior of the four-point function near $\tau=\t=0$. This behavior \eqref{eq:AsymptoticScalars} was fixed by demanding the existence of a non-trivial macroscopic limit. For non-scalar probes determining the macroscopic limit requires taking into account that the two insertion directions become aligned, namely
\begin{equation}
\vec n_3 \to \vec n_2,
\end{equation}
in other words
\begin{equation}
\vec n_3 = \vec n_2 + O(\t).
\end{equation}
Using \eqref{eq:SVgf}, \eqref{eq:VVgf}, \eqref{eq:SVfh}, and \eqref{eq:VVfh} we conclude that the functions have the following asymptotic behavior
\begin{align}
& h_2(\tau, \t), h_{\d}(\tau, \t), h_{32}(\tau, \t) \sim \f{1}{\tau^d}, \\
& h_3(\tau, \t), h_{22}(\tau, \t), h_{33}(\tau, \t) \sim \f{1}{\tau^{d+1}}, \\
& h_{23}(\tau, \t) \sim \f{1}{\tau^{d+2}}.
\end{align} 
Introducing the notation
\begin{align}
\m_{V,\ell} = \n^+_{\ell} \, \f{\ell+d-2}{\ell+d-2-\omega_\ell} ,
\end{align}
we can rewrite the equations as follows. For scalars
\begin{align}
h (\tau, x) & = e^{\tau} \cos\t 
+ \sum_{\ell=0}^\infty |\m_{S,\ell}|^2 \, e^{\omega_\ell \tau} \, C^{(d/2-1)}_\ell(\t),
\end{align}
for scalar-vector
\begin{align}
\label{eq:hSVConserved}
h_{2} (\tau, \t) & = e^{\tau} \cos \t  + \sum_{\ell=1}^\infty \m_{S,\ell} \, \m_{V,\ell}  \,  
e^{\omega_\ell \tau} \, C_{\ell}^{(d/2-1)}(\cos \t), \\
h_{3} (\tau, \t) & = \f{e^{\tau}}{d-1} + \f{d-2}{d-1} \sum_{\ell=1}^\infty \m_{S,\ell} \, \m_{V,\ell}  \, \f{\omega_\ell}{J_{d,\ell}^2} e^{\omega_\ell \tau} \, C_{\ell-1}^{(d/2)}(\cos \t),
\end{align}
and for vectors
\begin{align}
\label{eq:hVVConserved}
h_{23} (\tau, \t) & = \f{d(d-2)}{(d-1)^2} \sum_{\ell=2}^\infty
\l | \m_{V,\ell} \r |^2 \f{\omega_\ell^2}{J_{d,\ell}^4} \,
e^{\omega_\ell \tau} \, C_{\ell-2}^{(d/2+1)}(\cos \t), \\
h_{\d} (\tau, \t) & = \f{e^{\tau}}{(d-1)^2} + \f{d-2}{(d-1)^2} \sum_{\ell=1}^\infty
\l | \m_{V,\ell}  \r |^2 \f{\omega_\ell^2}{J_{d,\ell}^4} \,
e^{\omega_\ell \tau} \, C_{\ell-1}^{(d/2)}(\cos \t), \nn \\
h_{22} (\tau, \t) & = \f{e^{\tau}}{d-1} + \f{d-2}{d-1} \sum_{\ell=1}^\infty
\l | \m_{V,\ell} \r |^2 \, \f{\omega_\ell}{J_{d,\ell}^2} 
e^{\omega_\ell \tau} \, C_{\ell-1}^{(d/2)}(\cos \t), \nn \\
h_{33} (\tau, \t) & =  \f{e^{\tau}}{d-1} + \f{d-2}{d-1} \sum_{\ell=1}^\infty
\l | \m_{V,\ell}  \r |^2 \, \f{\omega_\ell}{J_{d,\ell}^2} 
e^{\omega_\ell \tau} \, C_{\ell-1}^{(d/2)}(\cos \t), \nn \\
h_{32} (\tau, \t) & = e^{\tau} \cos \t + \sum_{\ell=1}^\infty
\l | \m_{V,\ell} \r |^2 \,
e^{\omega_\ell \tau} \, C_{\ell}^{(d/2-1)}(\cos \t). \nn
\end{align}
Lower bounds for the sums are partially derived using the Ward identity which in this case
\begin{align}
\int_0^\pi h_2(\tau, \t) \sin^{d-2} \t d \t = \int_0^\pi h_{32}(\tau, \t) \sin^{d-2} \t d \t = 0.
\end{align}
Assuming $N$ Regge trajectories and introducing yet one more time new notation
\begin{align}
x_i & = \omega_{i,\ell}^2, \quad z = J_{d,\ell}^2=\f{\ell(\ell+d-2)}{d-1}, \\
A_i & = \f{d(d-2)}{2\ell+d-2} \f{|\m_{S,i,\ell}|^2}{\omega_{i,\ell}}, \\
B_i & = \f{d(d-2)}{2\ell+d-2} \f{|\m_{V,i,\ell}|^2}{\omega_{i,\ell}},
\end{align}
we obtain the following (independent) continuity equations (for $n\geq 1$)

\begin{equation}
\begin{array}{c|c|c}
\text{Function} & \text{Equation} & \ell \geq \quad \\
\hline
h(\tau, \t) & \dst\sum_{i=1} ^N A_i (z) x_i^n(z) = P^{(S)}_n (z) & 0 \\

h_{2}(\tau, \t) & \dst\sum_{i=1} ^N \sqrt{A_i (z) B_i(z)} \, x_i^n(z) = P^{(2)}_n (z) & 1 \\

h_{3}(\tau, \t) & \dst\sum_{i=1} ^N \sqrt{A_i (z) B_i(z)} \, x_i^n(z) = z P^{(3)}_{n-1}(z) & 1 \\
\end{array}
\end{equation}
and
\begin{equation}
\begin{array}{c|c|c}
\text{Function} & \text{Equation} & \ell \geq \quad \\
\hline

h_{\d}(\tau, \t) & \quad \dst\sum_{i=1} ^N B_i (z) x_i^{n+1}(z) = z^2 P^{(\delta)}_{n-1} (z) \quad & 1 \\

h_{22}(\tau, \t) & \quad \dst\sum_{i=1} ^N B_i (z) x_i^n(z) = z P^{(22)}_{n-1} (z) \quad & 1 \\

h_{32}(\tau, \t) & \quad \dst\sum_{i=1} ^N B_i (z) x_i^n(z) = P^{(32)}_n (z) \quad & 1
\end{array}
\end{equation}

For $\ell>1$ all polynomials $P^{(\a)}_n(z)$ satisfy the following order-$N$ recurrence (see \eqref{eq:RecurrenceScalar})
\begin{equation}
\label{eq:OrderNRecurrence}
P^{(\a)}_{N+n}(z) +\sum_{k=1}^{N} (-1)^k P^{(\a)}_{N+n-k}(z) Q_k(z)=0,
\end{equation}
and the spectrum $x_i(z)$ corresponds to the roots of the following equation with polynomial coefficients
\begin{equation}
x^N +\sum_{k=1}^{N} (-1)^k x^{N-k} Q_k(z)=0,
\end{equation}
Comparing equations for $h_{\d}$, $h_{22}$, and $h_{32}$
 we see that 
\begin{equation}
P^{(2)}_n (z) = z P^{(3)}_{n-1}(z), \quad n \geq 1.
\end{equation}
Similarly, equations for $h_{\d}$, $h_{22}$, and $h_{32}$ imply that 
\begin{equation}
\label{eq:P32Expression}
P^{(32)}_1 (z) = z P^{(22)}_{0}, \quad P^{(32)}_n (z) = z P^{(22)}_{n-1}(z) = z^2 P^{(\delta)}_{n-2}(z), \quad n\geq 2.
\end{equation}
As a result, there are three systems of Vandermond-like equations
\begin{align}
& \sum_{i=1} ^N A_i (z) x_i^n(z) = P^{(AA)}_n (z), \quad 1 \leq n \leq N, \\
& \sum_{i=1} ^N \sqrt{A_i (z) B_i(z)} \, x_i^n(z) = P^{(AB)}_n (z), \quad 1 \leq n \leq N, \\
& \sum_{i=1} ^N B_i (z) x_i^n(z) = P^{(BB)}_n (z), \quad 1 \leq n \leq N,
\end{align}
with
\begin{align}
P^{(AA)}_n (z) & = P^{(S)}_n (z), \quad n \geq 1\\
P^{(AB)}_n (z) & = z P^{(3)}_{n-1} (z), \quad n \geq 1 \\
P^{(BB)}_n (z) & = \l\{ 
\begin{array}{l}
z P^{(22)}_{0} (z), \quad n=1, \\
z^2 P_{n-2}^{(\delta)}(z), \quad n\geq 2.
\end{array}
\r.
\end{align}
The matrix $V_{ni} = x_i^n$ with $i,n = 1,\dots, N$, is invertible provided there are no zero modes and no degeneracy, $x_i \neq 0$ and $x_i\neq x_j$. Therefore, in this case the systems of equations viewed separately (treating $\sqrt{A_i B_i} =C_i$ as a separate variable) can always be solved. However, the consistency of the three imposes constraints on the possible choices of polynomials $P^{(IJ)}_n(z)$.

Equations for $\ell=0$ and $\ell=1$ should be solved separately. For instanceIn, for a non-degenerate case (see also \eqref{eq:ScalarNonDegenerateOmega} and \eqref{eq:ScalarNonDegenerateMu}), the descendant contribution, $\ell=1$, implies that
\begin{equation}
\label{eq:DescendantConstraint}
A_1(1)=B_1(1)=1, \quad x_1(1) = 1.
\end{equation}

\subsection{One Regge trajectory}

For $N=1$ we have
\begin{equation}
x=Q_1(z),
\end{equation}
therefore we obtain from $n=1$ equations
\begin{align}
\label{eq:ABN1Solution}
A(z) & = \f{P^{(S)}_1 (z)}{Q_1(z)} \\
\sqrt{A(z) B(z)} & = \f{z P^{(3)}_0}{Q_1(z)}, \\
B(z) & = \f{z P^{(22)}_{0}}{Q_1(z)},
\end{align}
which for $n\geq 2$ translate into
\begin{align}
P^{(S)}_n (z) & = P^{(S)}_1 (z) Q^{n-1}_1(z), \\
P^{(3)}_{n-1} (z) & = P^{(3)}_0 Q^{n-1}_1(z), \\
P^{(\d)}_{n-2}(z) & = P^{(22)}_{0} \f{Q^{n-1}_1(z)}{z}.
\end{align}
Since $P_{n}^{(\d)}(z)$ is a polynomial, it necessitates that $Q_{1}(z)$ be divisible by $z$. In other words,
\begin{equation}
Q_1(z) = q_{11} z.
\end{equation}
The consistency for the solution \eqref{eq:ABN1Solution} also demands that
\begin{equation}
P^{(S)}_1 (z) = z \f{\l [ P^{(3)}_0\r ]^2}{P^{(22)}_{0}},
\end{equation}
Eventually, the constraint from $\ell=1$ leads to
\begin{align}
q_{11} = P^{(3)}_{0} = P^{(22)}_{0} = 1.
\end{align}
As a result, the solution is given by
\begin{equation}
x = z, \quad A(z) =  B(z)  = 1,
\end{equation}
reproducing the EFT result given in \eqref{eq:hEFT}, \eqref{eq:h2EFT}, \eqref{eq:h3EFT}, and \eqref{eq:h23EFT}-\eqref{eq:h32EFT}.

\subsection{Two Regge trajectories}

For multiple Regge trajectories we derive one important constraint. It follows from \eqref{eq:OrderNRecurrence} for $(\a)=(32)$ and $n=1$
\begin{equation}
P^{(32)}_{N+1}(z) - P^{(32)}_{N}(z) Q_1(z) \dots + (-1)^{N-1} P^{(32)}_{2}(z) Q_{N-1}(z) + (-1)^{N} P^{(32)}_{1}(z) Q_{N}(z)= 0,
\end{equation}
which using \eqref{eq:P32Expression} becomes
\begin{equation}
z^2 \l [ P^{(\d)}_{N-1}(z) - P^{(\d)}_{N-2}(z) Q_1(z) \dots + (-1)^{N-1} P^{(\d)}_{0}(z) Q_{N-1}(z) \r ] + (-1)^{N} z P^{(22)}_{0}(z) Q_{N}(z)= 0.
\end{equation}
Due to unitarity $P^{(22)}_0(z)\neq 0$. Therefore, $Q_N(z)$ should be divisible by $z$. In other words
\begin{equation}
Q_N(0)=0.
\end{equation}
From the EFT perspective, this is a necessary condition for there to be a conserved current corresponding to the shifts of the Goldstone $\p_\m \pi$.

In particular, for two Regge trajectories we have
\begin{equation}
x_{1,2}(z) =\f{1}{2} \l [ Q_1(z) \mp \sqrt{Q_1^2(z) - 4Q_2(z)} \r ].
\end{equation}
Denoting $C_i(z) = \sqrt{A_i(z)B_i(z)}$ we have
\begin{align}
A_1 (z) & = \f{P^{(AA)}_2(z) - x_2(z) P^{(AA)}_1(z)}{x_{1}(z) \Big[ x_1(z)-x_2(z) \Big ]}, \quad 
A_2 (z) = \f{P^{(AA)}_2(z) - x_1(z) P^{(AA)}_1(z)}{x_{2}(z) \Big[ x_2(z)-x_1(z) \Big ]}, \\
B_1 (z) & = \f{P^{(BB)}_2(z) - x_2(z) P^{(BB)}_1(z)}{x_{1}(z) \Big[ x_1(z)-x_2(z) \Big ]}, \quad 
B_2 (z) = \f{P^{(BB)}_2(z) - x_1(z) P^{(BB)}_1(z)}{x_{2}(z) \Big[ x_2(z)-x_1(z) \Big ]}, \\
C_1 (z) & = \f{P^{(AB)}_2(z) - x_2(z) P^{(AB)}_1(z)}{x_{1}(z) \Big[ x_1(z)-x_2(z) \Big ]}, \quad 
C_2 (z) = \f{P^{(AB)}_2(z) - x_1(z) P^{(AB)}_1(z)}{x_{2}(z) \Big[ x_2(z)-x_1(z) \Big ]}.
\end{align}
The consistency constraints demand
\begin{align}
\Big[ P^{(AB)}_2(z) - x_2(z) P^{(AB)}_1(z) \Big ]^2 & = 
\Big[ P^{(AA)}_2(z) - x_2(z) P^{(AA)}_1(z) \Big ] \Big[ P^{(BB)}_2(z) - x_2(z) P^{(BB)}_1(z) \Big ], \\
\Big[ P^{(AB)}_2(z) - x_1(z) P^{(AB)}_1(z) \Big ]^2 & = 
\Big[ P^{(AA)}_2(z) - x_1(z) P^{(AA)}_1(z) \Big ] \Big[ P^{(BB)}_2(z) - x_1(z) P^{(BB)}_1(z) \Big ].
\end{align}
For a generic choice of polynomials $Q_i(z)$ these two constraints imply\footnote{These equations can be obtained by equating powers of $z$ and $x_i$ on both sides, treating $x_i$ as independent variables. The analysis should be modified for specific choices of polynomials $Q_i(z)$ resulting in an explicit dependence of $x_i$ on $z$.}
\begin{align}
P^{(AA)}_1(z) & = a_{11} z, \quad P^{(AB)}_1(z) = c_{11} z, \quad P^{(BB)}_1(z) = b_{11} z \\
P^{(AA)}_2(z) & = a_{22} z^2, \quad P^{(AB)}_2(z) = c_{22} z^2, \quad P^{(BB)}_2(z) = b_{22} z^2,
\end{align}
with 
\begin{align}
c_{11}=\sqrt{a_{11}b_{11}}, \quad c_{22}=\sqrt{a_{22}b_{22}}, \quad a_{11}b_{22}=a_{22}b_{11}.
\end{align}
In other words we only have three independent parameters.

The constraint \eqref{eq:DescendantConstraint} at $z=1$ demands that at least one root of the characteristic polynomial be $1$.
It is convenient therefore to parametrize the two polynomials $Q_1(z)$ and $Q_2(z)$ with the other root, which we denote $\b$
\begin{align}
Q_1(z) & = q_{11}(z-1) + 1+\b, \\
Q_2(z) & = q_{22}z(z-1) + \b z. 
\end{align}
As a result we obtain. For $\b >1$
\begin{align}
\label{eq:2ReggeBeta>1}
P^{(AA)}_1(z) & = P^{(AB)}_1(z) = P^{(BB)}_1(z) = b_{11} z, \\
P^{(AA)}_2(z) & = P^{(AB)}_2(z) = P^{(BB)}_2(z) = \Big [  1+ \b ( b_{11} -1 ) \Big ] z^2, \nn \\
A_2(1) & = B_2(1) = C_2(1) = \f{b_{11}-1}{\b}, \quad A_1(1)=1 \nn
\end{align}
for $\b <1$
\begin{align}
\label{eq:2ReggeBeta<1}
P^{(AA)}_1(z) & = P^{(AB)}_1(z) = P^{(BB)}_1(z) = b_{11} z, \\
P^{(AA)}_2(z) & = P^{(AB)}_2(z) = P^{(BB)}_2(z) = \Big [ \b ( \b -1 ) + b_{11} \Big ] z^2, \nn \\
A_1(1) & = B_1(1) = C_1(1) = b_{11}-\b, \quad A_2(1) = 1, \nn
\end{align}
and for $\b=1$ (degenerate case)
\begin{align}
\label{eq:2ReggeDegenerate}
P^{(AA)}_1(z) & = P^{(AB)}_1(z) = P^{(BB)}_1(z) = (1+b) z, \\
P^{(AA)}_2(z) & = P^{(AB)}_2(z) = P^{(BB)}_2(z) = (1+b) z^2, \nn \\
A_i(1) & = B_i(1), \quad B_1(1) + B_2(1) =  b + 1. \nn
\end{align}

\section{EFT realizations \label{sec:EFTRealizations}}

In this section we demonstrate that certain two-Regge-trajectory bootstrap solutions admit a realization within a local effective field theory at large charge. We first review the general structure of the large-charge EFT for the Goldstone mode only and later discuss additional light degrees of freedom.

\subsection{Setup}

The system is described by the following Lagrangian on the Euclidean cylinder $\mathbb R \times \mathbb S^{d-1}$
\begin{equation}
\mc L = - c_d \l [ - \l ( \p \c \r )^2 \r ]^{d/2} + c_{d-2} \mc R \l [ - \l ( \p \c \r )^2 \r ]^{d/2-1} + \dots
\end{equation}
The current is given by
\begin{equation}
\hat J^\m = -\f{\p \mc L}{\p \p_\m \c} = - c_d d \l [ - \l ( \p \c \r )^2 \r ]^{d/2-1} \p^\m \c.
\end{equation}
The classical solution with charge $Q$ is given by
\begin{equation}
\bar \c = - i \m \tau,
\end{equation}
leading to the following energy (scaling dimension) of the ground state
\begin{equation}
\label{eq:QmuR-Relation}
\D_Q = c_d (d-1) \Omega_{d-1} (\m R)^d,
\end{equation}
with the parameter $\m$ related to the charge $Q$ as
\begin{equation}
Q = c_d d \Omega_{d-1} (\m R)^{d-1}.
\end{equation}
The fluctuations $\tilde \pi$ around the classical background 
\begin{equation}
\c = \bar \c + \tilde \pi,
\end{equation}
in turn, result in the following quadratic action
\begin{equation}
\mc L = \f{c_d d (d-1)}{2}\m^{d-2} \l [ \dot {\tilde \pi}^2 + \f{1}{d-1} (\vec \nabla \tilde \pi)^2 \r].
\end{equation}
Rescaling the fields as
\begin{equation}
\pi = \sqrt{c_d d (d-1)} \m^{d/2-1} \tilde \pi,
\end{equation}
yields the canonically normalized quadratic Lagrangian
\begin{equation}
\mc L = \f{1}{2} \l [ \dot {\pi}^2 + \f{1}{d-1} (\vec \nabla \pi)^2 \r].
\end{equation}
Therefore, the Hilbert space of fluctuations has the Fock space structure corresponding to creation operators $a^\dagger_{\ell, \vec m}$ with energy $J_{d,\ell}$ and the corresponding degeneracy. The field can be represented as
\begin{equation}
\label{eq:piModeExpandsion}
\pi (\tau, \vec n) = \sum_{\ell, m} \f{1}{\sqrt{2J_{d,\ell}} R^{d/2-1}} \l [ a_{\ell m} Y_{\ell m} (\vec n) e^{-J_{d,\ell} \tau/R} + a^\dagger_{\ell m} Y^*_{\ell m} (\vec n) e^{J_{d,\ell} \tau/R} \r ].
\end{equation}
Any operator at low energies can be expressed in terms of the Goldstone modes by simply matching the corresponding quantum numbers: the scaling dimension and the $U(1)$ charge. For instance, a scalar operator with scaling dimension $\d$ and the charge $q$ can be written as
\begin{align}
\label{eq:EFT-operator-matching}
\mc O _q & =
 \l \{ C_0 \l [ - \l ( \p \c \r )^2 \r ]^{\d/2} + C_{2} \mc R \l [ - \l ( \p \c \r )^2 \r ]^{d/2-1} + \dots \r \} e^{i q \c} \\
& = 
\m^\d
\l \{ C_0 \l [ 1+\f{i\d \dot {\tilde \pi} }{\m} \r ] + C_{2} \m^{-2} \mc R \l [ 1+\f{i (\d-2) \dot {\tilde \pi} }{\m} \r ] + \dots \r \} e^{i q \c} \\
& = 
\label{eq:ScalarSaddle}
\m^\d
\l \{ C_0 \l [ 1+\f{i\d \dot {\pi} }{\sqrt{c_d d (d-1)} \m^{d/2}} \r ] + C_{2} \m^{-2} R \l [ 1+\f{i (\d-2) \dot {\pi} }{\sqrt{c_d d (d-1)} \m^{d/2}} \r ] + \dots \r \} e^{i q \c}.
\end{align}
Similarly, we can find the expression for the current in terms of the fluctuations
\begin{align}
\label{eq:CurrentEFT}
\hat J^\m & = i \f{Q}{\Omega_{d-1}R^{d-1}} \Bigg \{ \d^\m _ \tau + \f{i}{\m} \Big [ (d-2) \d^\m_\tau \dot {\tilde \pi} + \p^\m \tilde \pi \Big ] \Bigg \} \\
& = 
i \f{Q}{\Omega_{d-1}R^{d-1}} \Bigg \{ \d^\m _ \tau + \f{i}{\sqrt{c_d d (d-1)} \m^{d/2}} \Big [ (d-2) \d^\m_\tau \dot \pi + \p^\m \pi \Big ] \Bigg \}.
\end{align}

\subsection{EFT with additional light fields}

Here we only consider EFT like quadratic Lagrangians with two fields. We assume that the Lagrangian corresponds to a theory with a broken $U(1)$ symmetry. As a result, we expect the spectrum to contain at least one gapless (scalar) mode. We will discuss three scenarios: two-scalar, scalar-vector, and scalar-rank-two-tensor theories. The unbroken symmetry group on the cylinder corresponds to time translations and $SO(d-1)$ rotations.

\subsection{Two scalars}

The most general quadratic Lagrangian for two fields $\phi_1$, $\phi_2$ with at most two derivative can be written as
\begin{align}
\mc L = \f{A_{ab}}{2} \dot \phi^a \dot \phi^b - \f{B_{ab}}{2} \phi^a \dot \phi^b - \f{C_{ab}}{2} \nabla _i \phi^a \nabla ^i \phi^b - \f{D_{ab}}{2} \phi^a \phi^b.
\end{align}
We use orthogonal transformations to diagonalize $A_{ab}$ and then we perform field rescaling to set $A_{ab}=\d_{ab}$.
As a second step we choose to diagonalize the mass matrix $D_{ab}$. As a result, the Lagrangian becomes
\begin{align}
\mc L & = \f{1}{2} \dot \pi^2 + \f{1}{2} \dot r^2 - \f{m_r^2}{2}r^2 - \f{\gamma}{2} \l ( \pi \dot r - r \dot \pi \r ) \\ 
& \quad \quad - \f{c_{11}}{2} \nabla _i \pi \nabla ^i \pi - c_{12} \nabla _i \pi \nabla ^i r - \f{c_{22}}{2} \nabla _i r \nabla ^i r. \nn
\end{align}
The advantage of this parametrization is that it makes the shifts $\pi \to \pi + c$  manifestly a symmetry of the quadratic Lagrangian. The spectrum can be obtained from the following characteristic equation\footnote{As before
\begin{equation}
x = \omega^2, \quad z = J_{d,\ell}^2.
\end{equation}}
\begin{equation}
x^2 - x Q_1(z) + Q_2(z)=0,
\end{equation}
with polynomials
\begin{align}
Q_1 (z) & = (d-1)(c_{11}+c_{22})z + m^2 +\g^2, \\
Q_2 (z) & = (d-1)^2\l ( c_{11}c_{22} - c_{12}^2 \r ) z^2 + (d-1) c_{11} m^2 z.
\end{align}

\subsection{Scalar-vector}

The most general Lagrangian in this case
\begin{align}
\mc L & = \f{1}{2} \dot \pi^2 - \f{c_\pi^2}{2} \nabla _i \pi \nabla ^i \pi - \s \pi \nabla_j v^i + \g \dot \pi \nabla_j v^i \\
& \qquad + \f{1}{2} \dot v^i \dot v_i - 
\f{c_L^2}{2} \nabla _i v^i \nabla _j v^j - \f{c_T^2}{4} (\nabla _i v_j - \nabla _j v_i) (\nabla ^i v^j - \nabla^j v^i) - \f{m_v^2}{2} v_i v^i
\end{align}
Decomposing the vector into the longitudinal and the transverse components
\begin{equation}
v^i = \nabla^i \phi + u ^i, ~~ \nabla_i u^i = 0,
\end{equation}
results in
\begin{align}
\mc L & = \f{1}{2} \dot \pi^2 - \f{c_\pi^2}{2} \nabla _i \pi \nabla ^i \pi + \f{1}{2} \nabla_i \dot\phi \nabla_i \dot\phi -\f{c_L^2}{2} \nabla^2 \phi \nabla^2 \phi - \s \pi \nabla^2 \phi + \g \dot \pi \nabla^2 \phi - \f{m_v^2}{2} \nabla_i \phi \nabla^i \phi \nn \\
& \qquad + \f{1}{2} \dot u^i \dot u_i - \f{c_T^2}{4} (\nabla _i u_j - \nabla _j u_i) (\nabla ^i u^j - \nabla^j u^i) - \f{m_v^2}{2} u_i u^i.
\end{align}
The spectrum for the vector modes can be found using \eqref{eq:VectorEigenvalues}
\begin{equation}
\omega_{V,\ell}^2 = c_T^2 \l [  \ell(\ell+d-2) - 1 \r ] + \l [ m_v^2 + c_T^2 (d-2) \r ] 
= c_T^2 \l [ (d-1) J^2_{d,\ell} - 1 \r ] + \l [ m_v^2 + c_T^2 (d-2) \r ].
\end{equation}
To find the spectrum of scalar modes, we need to diagonalize the mixing. For $\ell=0$ there is no mixing and we get 
\begin{equation}
\omega_{\pi,0}^2 = 0.
\end{equation}
For $\ell\neq 0$, rescaling the field 
\begin{equation}
\phi \to \f{\sqrt{d-1}}{J_{d,\ell}},
\end{equation}
we obtain the same characteristic polynomial now with polynomials
\begin{align}
Q_1 (z) & = (d-1)(c_\pi^2 + c_L^2 + \g^2) z + m_v^2, \\
Q_2 (z) & = (d-1)^2c_\pi^2 c_L^2 z^2 + (d-1) (c_\pi^2 m_v^2 -\m^2) z.
\end{align}
Positivity and reality of the spectrum implies that
\begin{align}
c_\pi^2 \geq \f{\m^2}{m_v^2} \l ( 1 + \f{c_L^2 m_v^2}{\m^2+\g^2 m_v^2} \r ).
\end{align}

\subsection{Scalar-tensor}

For a scalar $\pi$ and a traceless symmetric tensor $h_{ij}$, the most general Lagrangian is given by
\begin{align}
\mc L & = \f{1}{2} \dot \pi^2 - \f{c_\pi^2}{2} \nabla _i \pi \nabla ^i \pi 
+ \g \pi \nabla_i \nabla_j h^{ij} \\
& \qquad + \f{1}{2} \dot h^{ij} \dot h_{ij} 
- \f{c_1^2}{2} \nabla _k h_{ij} \nabla^k h^{ij} - \f{c_2^2}{2} \nabla_i h^{ik} \nabla^j h_{jk} 
- \f{m_h^2}{2} h^{ij} h_{ij}.
\end{align}
The tensor $h_{ij}$ can be expanded into scalar, and divergence-free vector and tensor parts
\begin{equation}
h_{ij} = \l ( \nabla_i \nabla _ j - \f{g_{ij}}{d-1} \nabla^2 \r ) \phi + \nabla_i u_j + \nabla_j u_i + w_{ij}, ~~ \nabla_i u^i = 0, ~~ \nabla_i w^{ij} = 0.
\end{equation}
We only focus on the scalar component. Integrating by part we arrive at
\begin{align}
\mc L & = \f{1}{2} \dot \pi^2 - \f{c_\pi^2}{2} \pi \l (-\nabla ^2 \r ) \pi -
+  \f{d-2}{d-1} \g \pi \l [ (\nabla^2)^2 + (d-1) \nabla^2 \r ] \phi  \\
& \qquad 
+ \f{1}{2} \f{d-2}{d-1} \dot \phi \l [ (\nabla^2)^2 + (d-1) \nabla^2 \r ] \dot \phi
- \f{m_h^2}{2} \f{d-2}{d-1} \phi \l [ (\nabla^2)^2 + (d-1) \nabla^2 \r ] \phi \nn \\
& \qquad 
- \f{c_1^2}{2} \f{d-2}{d-1} \phi\l [ - (\nabla^2)^3 -3(d-1) (\nabla^2)^2 - 2 (d-1) \nabla^2 \r ]  \phi \\
& \qquad
- \f{c_2^2}{2} \l ( \f{d-2}{d-1} \r )^2 \phi \l [ - (\nabla^2)^3 - 2 (d-1) (\nabla^2)^2 - (d-1)^2 \nabla^2 \r ] \phi.  
\end{align}
Expanding in spherical harmonics
\begin{equation}
\pi (t, \vec n) = \sum_{\ell=0}^\infty \sum_{m} \pi_{\ell,m} (t) Y_{\ell,m}(\vec n), ~~ 
\phi (t, \vec n) = \sum_{\ell=2}^\infty \sum_m \f{\phi_{\ell,m} (t)}{\sqrt{(d-1)(d-2) J^2_{d,\ell} \l ( J^2_{d,\ell}- 1 \r )}} Y_{\ell,m}(\vec n),
\end{equation}
and integrating over the sphere $\mathbb S^{d-1}$ we get for $\ell \geq 2$
\begin{align}
L_{\ell,m} & = 
\f{1}{2} \dot \pi_{\ell,m}^2 - \f{c_\pi^2 (d-1)}{2} J_{d,\ell}^2 \pi_{\ell,m}^2 
+ \f{1}{2} \dot \phi_{\ell,m}^2 - \f{m_h^2}{2} \phi_{\ell,m}^2 \nn \\
& \qquad + \sqrt{(d-1)(d-2)} \g \pi_{\ell,m} \phi_{\ell,m} J_{d,\ell} \sqrt{J_{d,\ell}^2-1} \\
& \qquad 
- \f{1}{2} \phi_{\ell,m}^2 \Big [ c_1^2(d-1) (J_{d,\ell}^2-2) + c_2^2(d-2) (J_{d,\ell}^2-1) \Big ]. \nn
\end{align}
The polynomials in this case are given by
\begin{align}
x^2 - x \, Q_1 (z) + Q_2 (z) = 0, 
\end{align}
with
\begin{align}
Q_1(z) & = \Big [ c_\pi^2 (d-1) + c_1^2 (d-1) + c_2^2 (d-2) \Big ] z
+ \Big [ m_h^2 - 2 c_1^2 (d-1) -  c_2^2 (d-2) \Big ], \\
Q_2(z) & = (d-1) \bigg \{ c_\pi^2 \Big [ c_1^2 (d-1) + c_2^2 (d-2) \Big ] - (d-2) \g^2 \bigg \} z^2 \\
& \qquad - (d-1) \bigg \{ c_\pi^2 \Big [ 2 c_1^2 (d-1) + c_2^2 (d-2) \Big ] - (d-2) \g^2 \bigg \} z. \nn
\end{align}
In particular for
\begin{equation}
c_\pi^2 = \f{1}{d-1}, ~~ c_1^2 = \f{1}{d-1}, ~~ c_2 = 0, ~~ m_\pi^2 = 0, ~~ m_h^2 =2, ~~ 
\mu^2 = \f{a^2}{(d-1)(d-2)},
\end{equation}
we reproduce the spectrum obtained in~\cite{Jafferis:2017zna} with
\begin{equation}
Q_1(z) = 2z, ~~ Q_2(z) = (1-a^2)z^2 + a^2 z.
\end{equation}

\section{Conclusion \label{sec:Conclusion}}

A central question in the large-charge conformal bootstrap program is the status of effective field theory. At the order relevant for this work, EFT-like theories are quadratic theories of fluctuations around the large-charge state dictated by the spontaneously broken $U(1)$ symmetry, generically with additional light degrees of freedom.

The scalar-probe bootstrap of~\cite{Jafferis:2017zna} shows that crossing symmetry together with the existence of a non-trivial macroscopic limit leads to a parametrization of heavy-light CFT data (spectrum and OPE coefficients) in terms of polynomials $Q_n(z)$ and $P_n(z)$ with $n=1, \dots, N$, assuming that the spectrum is organized into a finite number $N$ of Regge trajectories. When only scalar probes are considered, the resulting solution space may appear broader than what is immediately recognized as arising from an EFT-like description. In this paper we considered vector probes, primarily the conserved $U(1)$ current, and found that they restrict the bootstrap solutions. Concretely, the bootstrap equations consistency translates into a nontrivial constraint on the polynomials
\begin{equation}
Q_N(0)=0\,.
\end{equation}
This condition is absent for purely scalar probes and is precisely what is required for the bootstrap spectrum to be compatible with the EFT expectations.

We further demonstrated the EFT realizations of certain two-Regge-trajectory bootstrap solutions. Starting from the most general local quadratic Lagrangian consistent with the symmetries we found generic spectra for theories with an additional light scalar, a vector, and a rank-two traceless symmetric tensor. In particular we demonstrate that the non-obviously pseudo particle interpretable two-Regge-trajectory spectrum found in~\cite{Jafferis:2017zna} can be obtained from a theory with an additional light rank-two traceless symmetric tensor. These results point into toward a tighter relation between the bootstrap and EFT-like descriptions at this order. 

To further clarify the relation at this order, it appears especially promising to analyze the two-Regge-trajectory case in more detail. A constructive approach could proceed in two steps. First, one should determine whether a given bootstrap spectrum obtained in Section~\ref{sec:Solutions} can be reproduced by a local quadratic Lagrangian with two fields. On the EFT side, this will likely require, extending the analysis beyond the scalar-tensor system to include higher-spin probes, while on the bootstrap side, it would be valuable to implement general unitarity/positivity constraints directly on the polynomial data, further restricting polynomials $P_{1,2}(z)$ and $Q_{1,2}(z)$. Second, incorporating tensor probes (most notably the stress tensor for imposing the local conformal symmetry) may provide further constraints beyond those coming from the $U(1)$ current. Finally, even for spectra that are reproducible within EFT, the bootstrap allows nontrivial freedom in the corresponding OPE coefficients (see Eqs.~\eqref{eq:2ReggeBeta>1}-\eqref{eq:2ReggeDegenerate}), and it remains to be understood in detail whether and how the parameter space of EFT realizations accommodates this freedom at the same order.

We emphasize that we do not claim that every algebraic bootstrap solution necessarily corresponds to a physical CFT, nor that it survives at higher orders in the large-$Q$ expansion. Likewise, we do not claim that every quadratic Lagrangian with a shift symmetry admits a consistent completion into a fully fledged UV-completable effective theory.
Nevertheless, establishing a sharp correspondence between bootstrap solutions and EFT realizations, or producing an explicit counterexample, would provide a decisive step toward clarifying the relation between the conformal bootstrap and effective field theory in large-charge sectors.

\section*{Acknowledgements}

We would like to thank Gabriel Cuomo for very useful discussions and suggestions. The work is supported by the National Science Foundation under Award No. 2310243.

\newpage
\appendix

\section{Details of vector bootstrap \label{app:VectorDetails}}

Here we present explicit computations. We first collect the relevant three-point data, then write the four-point tensor structures and their $s$-channel expansions, and finally organize the large-charge limit in a basis adapted to the macroscopic scaling and crossing symmetry.

\subsection{Three-point functions and OPE}

We start from the standard three-point structures \cite{Costa:2011mg,Costa:2011dw}. For two scalars and a spin-$\ell$ traceless symmetric tensor $T^{a_1\dots a_\ell}$ we have
\begin{equation}
\label{eq:3ptFunction-SSO}
\la \phi _ 3 (x_3)\, \phi_2(x_2)\, T^{a_1\dots a_\ell} (x_1) \ra
= \lambda_{\phi_3,\phi_2,\ell}\,
\f{V^{a_1}_{1,23}\dots V^{a_\ell}_{1,23} - \mathrm{traces}}{
x_{12}^{\D_1+\ell+\D_2-\D_3}\,
x_{23}^{\D_2+\D_3-\D_1-\ell}\,
x_{13}^{\D_1+\ell+\D_3-\D_2}},
\end{equation}
and for a scalar, a vector, and a spin-$\ell$ tensor
\begin{equation}
\label{eq:3ptFunction-SVO}
\la \phi _ 3(x_3)\, V^a_2(x_2)\, T_1^{a_1\dots a_\ell} (x_1)\ra =
\f{\alpha\, V^{a_1}_{1,23}\dots V^{a_\ell}_{1,23} V_{2,31}^a 
+ \frac{\beta}{\ell} \sum_{i=1}^{\ell} H_{12}^{a a_i}\,
V^{a_1}_{1,23}\dots \widehat{V^{a_i}_{1,23}}\dots V^{a_\ell}_{1,23}
- \mathrm{traces}}{
x_{12}^{\D_1+\ell+\D_2-\D_3}\,
x_{23}^{\D_2+\D_3-\D_1-\ell}\,
x_{13}^{\D_1+\ell+\D_3-\D_2}}.
\end{equation}
Here the hat indicates omission. Taking the limit
\begin{equation}
x_1 \to 0, ~~ x_3 \to \infty,
\end{equation}
in \eqref{eq:3ptFunction-SSO} and \eqref{eq:3ptFunction-SVO} and going to the cylinder using \eqref{eq:OperatorRescaled},
we obtain
\begin{equation}
\label{eq:SSTlambda}
\la \phi_3 | \tilde \phi_2(0,\vec n_2) | T^{a_1\dots a_\ell} \ra = \lambda_{3,2,\ell} S^{a_1\dots a_\ell} (\vec n_2),
\end{equation}
and 
\begin{align}
\label{eq:SJO-OPE}
\la \phi_3 | \tilde V^a_2 (0,\vec n_2) | T_1 \ra^{a_1\dots a_\ell} 
& = \bar \lambda^+_{3,2,\ell} S^{a a_1\dots a_\ell} (\vec n_2) \nn \\ 
& + \bar \lambda^-_{3,2,\ell} \l [ g^{a a_1} S^{a_2\dots a_\ell} (\vec n_2) + \dots +  g^{a a_\ell} S^{a_1\dots a_{\ell-1}} (\vec n_2) \r ] \\
& - \f{2 \bar \lambda^-_{3,2,\ell}}{d+2(\ell-2)}  \sum _{i < j}  g^{a_i a_j} S^{a a_1 \dots a_{i-1}a_{i+1} \dots a_{j-1}a_{j+1} \dots a_\ell} (\vec n_2) \nn
\end{align}
with
\begin{align}
\bar \lambda^+_{3,2,\ell} & = \a-2\b, \\
\bar \lambda^-_{3,2,\ell} & = \f{\b}{\ell}+\f{\a-2\b}{d+2(\ell-1)},
\end{align}
and where $S^{a_1\dots a_\ell} (\vec n)$ is the rank-$\ell$ traceless symmetric tensor
\begin{align}
S^{a_1\dots a_\ell} (\vec n_2) & = n^{a_1}\dots n^{a_\ell} - \mathrm{traces} \\
& = 
n^{a_1}\dots n^{a_\ell} - \f{1}{d+2(\ell-2)}\sum_{i <  j} g^{a_i a_j}n^{a_1} \dots n^{a_{i-1}} n^{a_{i+1}} \dots n^{a_{j-1}} n^{a_{j+1}} \dots n^{a_\ell}+\dots \nn
\end{align}

Similarly, we can derive
\begin{align}
\label{eq:OJS-OPE}
\la T^{a_1\dots a_\ell}_1 | \tilde V^a (\vec n_2) | \phi_3 \ra
& = \lambda^+_{3,2,\ell} S_{a a_1\dots a_\ell} (\vec n_2) \nn \\ 
& + \lambda^-_{3,2,\ell} \l [ g_{a a_1} S_{a_2\dots a_\ell} (\vec n_2) + \dots + g_{a a_\ell} S_{a_1\dots a_{\ell-1}} (\vec n_2) \r ] \\
& - \f{2 \lambda^-_{3,2,\ell}}{d+2(\ell-2)} \sum _{i < j}  g_{a_i a_j} S_{a a_1 \dots a_{i-1}a_{i+1} \dots a_{j-1}a_{j+1} \dots a_\ell} (\vec n_2).\nn
\end{align}
The two sets of couplings are related by
\begin{align} 
\label{eq:LambdaLambdabar1}
\l(
\begin{array}{c}
\bar \lambda^+_{3,2,\ell} \\
\bar \lambda^-_{3,2,\ell}
\end{array} \r)
 & = M
 \l(
\begin{array}{c}
 \lambda^+_{3,2,\ell} \\
 \lambda^-_{3,2,\ell}
\end{array} \r)^*,
\end{align}
with
\begin{align}
\label{eq:LambdaLambdabar2} 
M = M^{-1}
 & = \l( 
 \begin{array}{ccc}
 \dst - \f{d-2}{2\ell + d-2} && -2 \ell \\ \\
 \dst - \f{2(d+\ell-2)}{(2\ell + d-2)^2}  && \dst \f{d-2}{2\ell + d - 2}
 \end{array} \r).
\end{align}
If the vector $V^a$ corresponds to a conserved current $J^a$, the coefficients $\a$ and $\b$ are related as explained in~\cite{Costa:2011mg}
\begin{equation}
\f{\a}{\b} = \f{\D_3-\D_1+\ell+d-2}{\D_3-\D_1},
\end{equation}
implying that
\begin{equation}
\label{eq:LambdaRatio}
\f{\bar \lambda^-_{3,2,\ell}}{\bar \lambda^+_{3,2,\ell} } = \f{\l ( \D_3-\D_1+\ell \r )\l ( \ell +d-2 \r )}{\ell \l ( \D_1-\D_3+\ell +d-2 \r ) \l ( 2 \ell+d-2 \r )}, \qquad \f{\lambda^-_{3,2,\ell}}{\lambda^+_{3,2,\ell}} = \f{\l ( \D_1-\D_3+\ell \r )\l ( \ell +d-2 \r )}{\ell \l ( \D_3-\D_1+\ell +d-2 \r ) \l ( 2 \ell +d - 2 \r )}.
\end{equation}

\subsection{Four-point functions and $s$-channel expansions}

We now consider four-point functions with scalar external operators at $x_1$ and $x_4$, and scalar or vector probes at $x_2$ and $x_3$.

\paragraph{Scalar-vector.}

The four-point function involving one scalar and one vector probe can be written as
\begin{align}
\label{eq:4ptFunctionSSVS-generic}
\la \phi_4 \phi_3 V_2^{a} \phi_1 \ra &
\equiv \la \phi_4(x_4) \phi_3 (x_3) J^{a} (x_2) \phi_1(x_1) \ra \\
& = \f{F_4 (u,v) V_{2,31}^a + F_1 (u,v) V_{2,34}^a}{x_{12} ^{\D_1+\D_2+1} x_{34} ^{\D_3+\D_4}}
\l( \f{x_{24}}{x_{14}}\r )^{\D_1-\D_2-1}\l( \f{x_{14}}{x_{13}}\r )^{\D_3-\D_4}. \nn
\end{align}
In the limit $x_1\to 0$ and $x_4\to \infty$ we obtain 
\begin{align}
\label{eq:4ptScalarVectorS-channelOPE}
n_2^a G_2 (u,v) +n_3^a G_3 (u,v) & = \la \phi_4| \tilde \phi_3 (\vec n_3) e^{-H(\tau_3-\tau_2)} \tilde J^a (\vec n_2) | \phi_1 \ra \\ 
&= \sum_E |z|^{E} \la \phi_4 | \tilde \phi_3 (\vec n_3) | E \ra \la E | \tilde V^a (\vec n_2) | \phi_1 \ra, \nn
\end{align}
where we introduced the functions $G_i(u,v)$ as linear combinations of $F_i(u,v)$
\begin{align}
G_2 (u,v) & = - \l [ F_1 (u,v) + (u-v) F_4(u,v) \r ], \\
G_3 (u,v) & = \f{F_1(u,v) + u F_4 (u,v) }{\sqrt{u}}.
\end{align}

\paragraph{Vectors.}
For two vector probes we have
\begin{align}
\label{eq:4ptFunctionSVVS-generic}
\la \phi_4 V_3^a V_2^{b} \phi_1 \ra &
\equiv \la \phi_4(x_4) J^a_3 (x_3) J^{b} (x_2) \phi_1(x_1) \ra \\
& = \f{1}{x_{12} ^{\D_1+\D_2+1} x_{34} ^{\D_3+\D_4+1}}
\l( \f{x_{24}}{x_{14}}\r )^{\D_1-\D_2-1}\l( \f{x_{14}}{x_{13}}\r )^{\D_3-\D_4+1} \nn \\
& \qquad \Big [ x_{23}^2 H_{23}^{ab} F_{23} + F_{11} (u,v) V^a_{3,24}V_{2,34}^b + F_{14} (u,v) V^a_{3,24}V_{2,31}^b \nn \\
& \qquad \qquad + F_{41} (u,v) V^a_{3,21}V_{2,34}^b + F_{44} (u,v) V^a_{3,21}V_{2,31}^b \Big ]. \nn
\end{align}
In the limit $x_1\to 0$ and $x_4\to \infty$ it becomes
\begin{align}
\label{eq:4ptVectorVectorS-channelOPE}
&n_3^a n_3^b \, G_{33}(u,v) + n_3^a n_2^b \, G_{32} (u,v) 
+ n_2^a n_3^b \, G_{23} (u,v) + n_2^a n_2^b \, G_{22} (u,v) 
+\d^{ab} G_{\d} (u,v)
\nn \\
& ~~~~ = \la \phi_4| \tilde V_3^a (\vec n_3) e^{-H(\tau_3-\tau_2)} \tilde V_2^b (\vec n_2) | \phi_1 \ra = 
\sum_E |z|^{E} \la \phi_4 | \tilde V^a_3 (\vec n_3) | E \ra \la E | \tilde V^b_2 (\vec n_2) | \phi_1 \ra,
\end{align}
with
\begin{align}
G_{33} & = - u^{-3/2} \l [ u \l ( F_{11} + u F_{14} + 2 F_{23} \r ) + (1-v) \l ( F_{41}+u F_{44} \r ) \r ], \\
G_{32} & = F_{11} + (u-v)F_{14} + 2 F_{23} + (1-v) u^{-1} \l [ F_{41} + (u-v) F_{44}\r ], \\
G_{23} & = F_{11} + u F_{14} + 2 F_{23} + \f{F_{41}}{u}+F_{44}, \\
G_{22} & = - u^{-1/2} \l [ u \l ( F_{11} + (u-v) F_{14} + 2 F_{23} \r ) + \l ( F_{41} + (u-v) F_{44} \r ) \r ], \\
G_{\d} & = v u^{-1/2} F_{23}.
\end{align}

There is yet another basis
\begin{align}
u^{\f{\D_1+\D_4}{4}}H_{23} & = G_{23}, \quad u^{\f{\D_1+\D_4}{4}}H_{\d} = G_{\d}, \\
u^{\f{\D_1+\D_4}{4}}H_{22}& = G_{22}+G_{23} \f{1+u-v}{2\sqrt{u}} + G_\d, \\
-u^{\f{\D_1+\D_4}{4}}H_{33}& = G_{33}+G_{23} \f{1+u-v}{2\sqrt{u}} + G_\d, \\
-u^{\f{\D_1+\D_4}{4}}H_{32}& = G_{32} + ( G_{22} + G_{33} + G_\d ) \f{1+u-v}{2\sqrt{u}} + G_{23} \l ( \f{1+u-v}{2\sqrt{u}} \r )^2.
\end{align}

\subsection{Conformal block expansion and combined $s$-channel result}

For setting up the bootstrap equations we need the contributions to the four-point functions from primary operators of spin $\ell$ and from the first descendant corresponding to the exchange of a scalar.

\subsubsection{Descendants}

For pure scalars, the contribution of a scalar of dimension $\D$ and its first descendant is
\begin{align}
(1')_{SS} & = \lambda_{\phi_4,\phi_3,\D} (\lambda_{\phi_1,\phi_2,\D})^* \l [1 + 2 \D |z| a_{34}^S a_{21}^S \, \vec n_2 \vec n_3 \r ].
\end{align}
For scalar-vector probes, the scalar primary plus descendant contribution is
\begin{align}
(1')_{SV} & = \lambda_{\phi_4,\phi_3,\D} (\lambda_{\phi_1,V_2,\D})^* \l [ n_2^a \l (1 + 
2\D |z| a_{34}^S a_{21}^V \, \vec n_2 \vec n_3 \r ) + 2\D a_{34}^S b_{21}^V |z| n_3^a \r ],
\end{align}
and for vector-vector probes
\begin{align}
(1')_{VV} & = \lambda_{\phi_4,V_3,\D} (\lambda_{\phi_1,V_2,\D})^* \Bigg \{ 
-n_3^a n_2^b \l [ 1 + 2 \D |z| \l ( a_{34}^V a_{21}^V + 2 b_{34}^V a_{21}^V \r ) \, \vec n_2 \vec n_3 \r ] \nn \\
& ~~~~ + 2 \D |z| \Bigg [ b_{34}^V a_{21}^V n_2^a n_2^b - \l( a_{34}^V b_{21}^V + 2 b_{34}^V b_{21}^V \r ) n_3^a n_3^b + b_{34}^V b_{21}^V \d^{ab} \Bigg ] \Bigg \},
\end{align}
with the following OPE coefficients
\begin{align}
a_{ij}^S & = \f{\D_i-\D_j+\D}{2\D}, ~~ a_{ij}^V = \f{\D_i-\D_j+\D-1}{2\D}, ~~ b_{ij}^V = \f{1}{2\D}.
\end{align}

\subsubsection{Primaries}

\paragraph{Scalars.}
We start from the four-point function of scalars and use the representation \eqref{eq:4ptScalarS-channelOPE} obtained in the scalar section. The contribution from a primary state with spin $\ell$, namely from $| E \ra _{a_1\dots a_\ell}$, to the expansion
\begin{align}
(\ell)_{SS}= \la \phi_4 | \tilde \phi_3(\vec n_3) | T^{(\ell)} \ra \la T^{(\ell)} | \tilde \phi_2 (\vec n_2) | \phi_1 \ra
\end{align}
can be found using \eqref{eq:SSTlambda} and
\begin{align}
\label{eq:SSproduct}
S_{a_1\dots a_\ell} (\vec n_3) S_{a_1\dots a_\ell} (\vec n_2) & = \f{\ell!}{2^\ell \l(\f{d}{2}-1 \r )_\ell} C^{(d/2-1)}_\ell(\vec n_2 \vec n_3),
\end{align}
with the Gegenbauer polynomial $C^{(d/2-1)}_\ell(x)$. As a result, we obtain
\begin{align}
(\ell)_{SS} = \lambda_{4,3,\ell} \l ( \lambda_{1,2,\ell} \r )^* \, \f{\ell!}{2^\ell \l(\f{d}{2}-1 \r )_\ell} \, C^{(d/2-1)}_\ell(\vec n_2 \vec n_3)
\end{align}

\paragraph{Scalar-vector.}
In this case, we want to compute
\begin{align}
(\ell)_{SV} = \la \phi_4 | \tilde \phi_3(\vec n_3) | T \ra^{(\ell)}{}^{(\ell)}\la T | \tilde V^a (\vec n_2) | \phi_1 \ra.
\end{align}
Using
\begin{equation}
S_{a_1\dots a_\ell} (\vec x_3)  S_{a a_1\dots a_\ell} (\vec x_2) = \f{1}{\ell +1} \f{\p}{\p x_3^a} S_{a_1\dots a_{\ell+1}} (\vec x_3) S_{a_1\dots a_{\ell+1}} (\vec x_2),
\end{equation}
together with\footnote{This expression can be obtained by applying the expression
\begin{equation}
\f{\p}{\p x^a} = n^a\f{\p}{\p x} + \f{\p n^c}{\p x^a} \f{\p}{\p n^c},
\end{equation}
to $x^b$ leading to
\begin{equation}
\d^{ab} = \f{\p x^b}{\p x^a} = n^a n^b + \f{\p n^c}{\p x^a} x \d^{bc}.
\end{equation}}
\begin{equation}
\f{\p}{\p x^a} = n^a\f{\p}{\p x} + \f{1}{x} \l ( \d^{ab} -n^a n^b \r ) \f{\p}{\p n^b},
\end{equation}
and
\begin{equation}
S_{a_1\dots a_\ell} (\vec x) = x^\ell S_{a_1\dots a_\ell} (\vec n),
\end{equation}
we find
\begin{align}
S_{a_1\dots a_\ell} (\vec n_3)  S_{a a_1\dots a_\ell} (\vec n_2) & = (\ell+1)^{-1} x_2^{-\ell-1}x^{-\ell}_3 \l ( n_3^a\f{\p}{\p x_3} + \f{1}{x_3} \l ( \d^{ab} -n^a_3 n^b_3 \r ) \f{\p}{\p n^b_3} \r ) \l [ x_2^{\ell+1}x^{\ell+1}_3 G^{(d/2-1)}_{\ell+1}(\vec n_2 \vec n_3) \r ] \nn \\ & = n_3^a G^{(d/2-1)}_{\ell+1}(\vec n_2 \vec n_3) + (\ell+1)^{-1} \l [ n_2^a-n_3^a (\vec n_2 \vec n_3) \r ]\l ( G^{(d/2-1)}_{\ell+1}(\vec n_2 \vec n_3) \r )'.
\end{align}
Using the Gegenbauer identities
\begin{align}
\label{eq:GegenbauerRecursion1}
\f{d}{dx}C_\ell^{(\n)}(x) & = 2 \n C_{\ell-1}^{(\n+1)}(x), \\
\label{eq:GegenbauerRecursion2}
(\ell+\n) C_\ell^{(\n)}(x) & = \n \l [ C_{\ell}^{(\n+1)}(x) - C_{\ell-2}^{(\n+1)}(x) \r ], \\
\label{eq:GegenbauerRecursion3}
2 x (\ell+\n) C_\ell^{(\n)}(x) & = (\ell+1) C_{\ell+1}^{(\n)}(x) + (\ell+2\n-1) C_{\ell-1}^{(\n)}(x) 
\end{align}
we can show that
\begin{align}
S_{a_1\dots a_\ell} (\vec n_3)  S_{a a_1\dots a_\ell} (\vec n_2) & = 
\f{\ell!}{2^\ell \l ( \f{d}{2} \r )_\ell}
\l [ n_2 ^a C_{\ell}^{(d/2)}(\vec n_2 \vec n_3) - n_3 ^ a C_{\ell-1}^{(d/2)}(\vec n_2 \vec n_3) \r ].
\end{align}
As a result, using \eqref{eq:3ptFunction-SSO} and \eqref{eq:SJO-OPE} we obtain the following contribution from a spin-$\ell$ primary
\begin{align}
\label{eq:Spin-ell-OPE-contribution}
(\ell)_{SV}
& = 
\lambda_{4,3,\ell} \l ( d-2 \r ) \f{\ell!}{2^{\ell} \l(\f{d}{2} -1 \r )_\ell}
\l \{ n_2^a \l [
\f{\lambda^+_{3,2,\ell}}{2 \ell+ d - 2} C_{\ell}^{(d/2)}(\vec n_2 \vec n_3) - \lambda^-_{3,2,\ell} C_{\ell-2}^{(d/2)}(\vec n_2 \vec n_3) \r ] \r. \nn \\
& \quad \quad \quad \quad \quad \quad
- \l. n_3 ^a \l [ \f{\lambda^+_{3,2,\ell}}{2 \ell+ d - 2} - \lambda^-_{3,2,\ell} \r ] C_{\ell-1}^{(d/2)}(\vec n_2 \vec n_3) \r \}.
\end{align}

\paragraph{Vectors.}

For two vector probes, a fully pedestrian derivation becomes rather cumbersome, and it is convenient to use the machinery developed in~\cite{Costa:2011dw}. Here we only present the final result (compare with \eqref{eq:4ptVectorVectorS-channelOPE})
\begin{align}
(\ell)_{VV} & = \la \phi_4 | \tilde V^a_3(\vec n_3) | T \ra^{(\ell)}{}^{(\ell)}\la T | \tilde V_2^b (\vec n_2) | \phi_1 \ra \\
& = d (d-2) \f{\ell!}{2^{\ell} \l(\f{d}{2} -1 \r )_\ell} \Big [ n_3^a n_3^b \, c_{33,\ell} + n_3^a n_2^b \, c_{32,\ell} 
+ n_2^a n_3^b \, c_{23,\ell} + n_2^a n_2^b \, c_{22,\ell} +\d^{ab} c_{\d,\ell} \Big ], \nn
\end{align}
with (see \eqref{eq:LambdaLambdabar2})
\begin{align}
c_{\d,\ell} (\lambda^+, \lambda^-)& =  
\f{1}{d} \l [ \f{\lambda^+_{3,2,\ell}}{2 \ell+ d - 2} - \lambda^-_{3,2,\ell} \r ] 
\l [ \f{\bar \lambda^+_{3,2,\ell}}{2 \ell+ d - 2} - \bar \lambda^-_{3,2,\ell} \r ] C_{\ell-1}^{(d/2)} (\vec n_2 \vec n_3) \\
c_{23,\ell} (\lambda^+, \lambda^-)& = 
\l [ \f{\lambda^+_{3,2,\ell}}{2 \ell+ d - 2} - \lambda^-_{3,2,\ell} \r ] 
\l [ \f{\bar \lambda^+_{3,2,\ell}}{2 \ell+ d - 2} - \bar \lambda^-_{3,2,\ell} \r ] C_{\ell-2}^{(d/2+1)} (\vec n_2 \vec n_3), \\
c_{22,\ell} (\lambda^+, \lambda^-)& = 
-\l [ \f{\bar \lambda^+_{3,2,\ell} }{2 \ell+ d - 2} - \bar \lambda^-_{3,2,\ell} \r ]
\l [
\f{\lambda^+_{3,2,\ell}}{2 \ell+ d - 2} C_{\ell-1}^{(d/2+1)}(\vec n_2 \vec n_3) - \lambda^-_{3,2,\ell} C_{\ell-3}^{(d/2+1)}(\vec n_2 \vec n_3) \r ], \\
c_{33,\ell} (\lambda^+, \lambda^-)& = -\l [ \f{\lambda^+_{3,2,\ell} }{2 \ell+ d - 2} - \lambda^-_{3,2,\ell} \r ]
\l [
\f{\bar \lambda^+_{3,2,\ell}}{2 \ell+ d - 2} C_{\ell-1}^{(d/2+1)}(\vec n_2 \vec n_3) - \bar \lambda^-_{3,2,\ell} C_{\ell-3}^{(d/2+1)}(\vec n_2 \vec n_3) \r ]
\end{align}
and
\begin{align}
c_{32,\ell} (\lambda^+, \lambda^-)& = 
-\f{\lambda^+_{3,2,\ell} \bar \lambda^+_{3,2,\ell}}{(2\ell+d)(2 \ell+ d - 2)} \, C_{\ell}^{(d/2+1)} (\vec n_2 \vec n_3) - 
\lambda^-_{3,2,\ell} \bar \lambda^-_{3,2,\ell} \, \f{2\ell+d-2}{2\ell+d-4} \, C_{\ell-4}^{(d/2+1)} (\vec n_2 \vec n_3) \\
& \quad + \l [ 
\f{2\lambda^+_{3,2,\ell} \bar \lambda^+_{3,2,\ell}}{(2\ell+d)(2 \ell+ d - 2)^2} 
- \f{\lambda^+_{3,2,\ell} \bar \lambda^-_{3,2,\ell}+\lambda^-_{3,2,\ell} \bar \lambda^+_{3,2,\ell}}{2 \ell+ d - 2} - \f{2 \lambda^-_{3,2,\ell} \bar \lambda^-_{3,2,\ell} }{2 \ell+ d - 4}
\r ] C_{\ell-2}^{(d/2+1)} (\vec n_2 \vec n_3). \nn
\end{align}
Equivalently
\begin{align}
c_{\d,\ell} (\lambda^+, \lambda^-)& =  
\f{1}{d} \l | \f{\lambda^+_{3,2,\ell}}{2 \ell+ d - 2} - \lambda^-_{3,2,\ell} \r |^2 C_{\ell-1}^{(d/2)} (\vec n_2 \vec n_3) \\
c_{23,\ell} (\lambda^+, \lambda^-)& = 
\l | \f{\lambda^+_{3,2,\ell}}{2 \ell+ d - 2} - \lambda^-_{3,2,\ell} \r |^2 C_{\ell-2}^{(d/2+1)} (\vec n_2 \vec n_3), \\
c_{22,\ell} (\lambda^+, \lambda^-)& = 
-\l [ \f{(\lambda^+_{3,2,\ell})^* }{2 \ell+ d - 2} - (\lambda^-_{3,2,\ell})^* \r ]
\l [
\f{\lambda^+_{3,2,\ell}}{2 \ell+ d - 2} C_{\ell-1}^{(d/2+1)}(\vec n_2 \vec n_3) - \lambda^-_{3,2,\ell} C_{\ell-3}^{(d/2+1)}(\vec n_2 \vec n_3) \r ], \\
c_{33,\ell} (\lambda^+, \lambda^-)& = \f{1}{2 \ell+ d - 2} \l ( \f{\lambda^+_{3,2,\ell}}{2 \ell+ d - 2} - \lambda^-_{3,2,\ell} \r ) \\ 
& ~~ \times
\l \{ \l [ \f{(d-2)(\lambda^+_{3,2,\ell})^*}{2 \ell+ d - 2} + 2\ell (\lambda^-_{3,2,\ell})^*\r ] C_{\ell-1}^{(d/2+1)} (\vec n_2 \vec n_3) \r. \nn \\
& \qquad \qquad - \l. \l [ \f{2(\ell+d-2)(\lambda^+_{3,2,\ell})^*}{2 \ell+ d - 2} - (d-2) (\lambda^-_{3,2,\ell})^*  \r ] C_{\ell-3}^{(d/2+1)} (\vec n_2 \vec n_3) \r \} \nn
\end{align}
and
\begin{align}
c_{32,\ell} (\lambda^+, \lambda^-)& = 
-\f{\lambda^+_{3,2,\ell}}{(2\ell+d)(2 \ell+ d - 2)} \, \l [ \f{(d-2)(\lambda^+_{3,2,\ell})^*}{2 \ell+ d - 2} + 2\ell (\lambda^+_{3,2,\ell})^* \r ] C_{\ell}^{(d/2+1)} (\vec n_2 \vec n_3) \\
& \quad + \l [ 
\f{2(\ell+d-1)|\lambda^+_{3,2,\ell}|^2}{(2\ell+d)(2 \ell+ d - 2)^2} 
- \f{d \, \lambda^+_{3,2,\ell} (\lambda^-_{3,2,\ell})^*}{(2\ell+d)(2 \ell+ d - 2)} \r. \\
& \qquad \qquad
+ \l. \f{d \, (\lambda^+_{3,2,\ell})^* \lambda^-_{3,2,\ell}}{(2 \ell+ d - 2)(2 \ell+ d - 4)} 
+ \f{2(\ell-1) |\lambda^-_{3,2,\ell}|^2}{2 \ell+ d - 4}
\r ] C_{\ell-2}^{(d/2+1)} (\vec n_2 \vec n_3) \nn \\
& \quad - \f{\lambda^-_{3,2,\ell}}{2\ell+d-4} \l [ \f{2(\ell+d-2)(\lambda^+_{3,2,\ell})^*}{2 \ell+ d - 2} - (d-2) (\lambda^+_{3,2,\ell})^* \r ] C_{\ell-4}^{(d/2+1)} (\vec n_2 \vec n_3). \nn
\end{align}

\subsection{Combined contribution}

Putting everything together, we obtain the following explicit expansions.

For scalars
\begin{align}
\label{eq:ExplicitExpansionScalars}
\la \phi_4| \tilde \phi_3 (\vec n_3) e^{-H(\tau_3-\tau_2)} \tilde \phi (\vec n_2) | \phi_1 \ra & \supset 
\lambda_{\phi_4,\phi_3,\D} (\lambda_{\phi_1,\phi_2,\D})^* |z|^\D \Big [1 + 2 \D |z| a_{34}^S a_{21}^S \, \vec n_2 \vec n_3 \Big ] \nn\\
& \quad \quad + \sum_\ell \f{\ell!}{2^{\ell} \l(\f{d}{2} -1 \r )_\ell} \lambda_{4,3,\ell} \l ( \lambda_{1,2,\ell} \r )^* |z|^{\D_\ell}\, 
C^{(d/2-1)}_\ell(\vec n_2 \vec n_3).
\end{align}

For scalar-vector
\begin{align}
\label{eq:ExplicitExpansionScalar-Vector}
& \la \phi_4| \tilde \phi_3 (\vec n_3) e^{-H(\tau_3-\tau_2)} \tilde V^a (\vec n_2) | \phi_1 \ra \\ 
& \supset 
\lambda_{\phi_4,\phi_3,\D} (\lambda_{\phi_1,V_2,\D})^*  |z|^\D \l [ n_2^a \l (1 + 
2\D |z| a_{34}^S a_{21}^V \, \vec n_2 \vec n_3 \r ) + 2\D a_{34}^S b_{21}^V |z| n_3^a \r ] \nn \\
& \quad \quad 
+ \l ( d-2 \r ) \sum_\ell \f{\ell!}{2^{\ell} \l(\f{d}{2} -1 \r )_\ell} \,
 \lambda_{4,3,\ell} |z|^{\D_\ell} \l \{ n_2^a \l [
\f{\lambda^+_{3,2,\ell}}{2 \ell+ d - 2} C_{\ell}^{(d/2)}(\vec n_2 \vec n_3) - \lambda^-_{3,2,\ell} C_{\ell-2}^{(d/2)}(\vec n_2 \vec n_3) \r ] \r. \nn \\
& \quad \quad \quad \quad \quad \quad
- \l. n_3 ^a \l [ \f{\lambda^+_{3,2,\ell}}{2 \ell+ d - 2} - \lambda^-_{3,2,\ell} \r ] C_{\ell-1}^{(d/2)}(\vec n_2 \vec n_3) \r \}, \nn
\end{align}
and for vectors
\begin{align}
\label{eq:ExplicitExpansionVectors}
&\la \phi_4| \tilde V_3^a (\vec n_3) e^{-H(\tau_3-\tau_2)} \tilde V_2^b (\vec n_2) | \phi_1 \ra \\
& =  \lambda_{\phi_4,V_3,\D} (\lambda_{\phi_1,V_2,\D})^* |z|^\D \Bigg \{ 
-n_3^a n_2^b \Big [ 1 + 2 \D |z| \l ( a_{34}^V a_{21}^V + 2 b_{34}^V a_{21}^V \r ) \, \vec n_2 \vec n_3 \Big ] \nn \\
& ~~~~ + 2 \D |z| \Bigg [ b_{34}^V a_{21}^V n_2^a n_2^b - \l( a_{34}^V b_{21}^V + 2 b_{34}^V b_{21}^V \r ) n_3^a n_3^b + b_{34}^V b_{21}^V \d^{ab} \Bigg ] \Bigg \} \nn \\
& \quad \quad + d (d-2) \sum_\ell \f{\ell!}{2^{\ell} \l(\f{d}{2} -1 \r )_\ell} |z|^{\D_\ell} \l ( n_3^a n_3^b \, c_{33,\ell} + n_3^a n_2^b \, c_{32,\ell} 
+ n_2^a n_3^b \, c_{23,\ell} + n_2^a n_2^b \, c_{22,\ell} +\d^{ab} c_{\d,\ell} \r ). \nn
\end{align}

\subsection{Crossing equations}

The crossing equations for scalar-vector probes become
\begin{align}
F^{(u)}_1 \l ( \f{1}{u}, \f{v}{u} \r ) & = F^{(s)}_4 (u,v)  u^{-\f{\D_1+\D_4}{2}}, \\
F^{(u)}_4 \l ( \f{1}{u}, \f{v}{u} \r ) & = F^{(s)}_1 (u,v)  u^{-\f{\D_1+\D_4}{2}},
\end{align}
equivalently
\begin{align}
G^{(u)}_2 \l ( \f{1}{u}, \f{v}{u} \r ) & = - u^{-\f{\D_1+\D_4}{2}} \l [ G^{(s)}_2 (u,v) + \f{1+u-v}{\sqrt{u}} G^{(s)}_3 (u,v)\r ], \\
G^{(u)}_3 \l ( \f{1}{u}, \f{v}{u} \r ) & = u^{-\f{\D_1+\D_4}{2}} G^{(s)}_3 (u,v),
\end{align}
and vectors probes
\begin{align}
F^{(u)}_{11} \l ( \f{1}{u}, \f{v}{u} \r ) & = u^{-\f{\D_1+\D_4}{2}} F^{(s)}_{44} (u,v), \\
F^{(u)}_{44} \l ( \f{1}{u}, \f{v}{u} \r ) & = u^{-\f{\D_1+\D_4}{2}} F^{(s)}_{11} (u,v), \\
F^{(u)}_{14} \l ( \f{1}{u}, \f{v}{u} \r ) & = u^{-\f{\D_1+\D_4}{2}} F^{(s)}_{41} (u,v), \\
F^{(u)}_{41} \l ( \f{1}{u}, \f{v}{u} \r ) & = u^{-\f{\D_1+\D_4}{2}} F^{(s)}_{14} (u,v), \\
F^{(u)}_{23} \l ( \f{1}{u}, \f{v}{u} \r ) & = u^{-\f{\D_1+\D_4}{2}} F^{(s)}_{23} (u,v),
\end{align}
equivalently
\begin{align}
u^{\f{\D_1+\D_4}{2}} G^{(u)}_\d \l ( \f{1}{u}, \f{v}{u} \r ) & = G^{(s)}_\d (u,v), \\
u^{\f{\D_1+\D_4}{2}} G^{(u)}_{23} \l ( \f{1}{u}, \f{v}{u} \r ) & = G^{(s)}_{23} (u,v), \\
u^{\f{\D_1+\D_4}{2}} G^{(u)}_{22} \l ( \f{1}{u}, \f{v}{u} \r ) & = - \l [ G^{(s)}_{22} (u,v) + \f{1+u-v}{\sqrt{u}} G^{(s)}_{23} (u,v) + 2 G^{(s)}_{\d} (u,v) \r ], \\
u^{\f{\D_1+\D_4}{2}} G^{(u)}_{33} \l ( \f{1}{u}, \f{v}{u} \r ) & = - \l [ G^{(s)}_{33} (u,v) + \f{1+u-v}{\sqrt{u}} G^{(s)}_{23} (u,v) + 2 G^{(s)}_{\d} (u,v) \r ], \\
u^{\f{\D_1+\D_4}{2}} G^{(u)}_{32} \l ( \f{1}{u}, \f{v}{u} \r ) & = G^{(s)}_{32} (u,v) + \f{(1+u-v)^2}{u} G^{(s)}_{23} (u,v) \nn \\
& \quad \quad + \f{1+u-v}{\sqrt{u}} \l [ 2 G^{(s)}_{\d} (u,v) + G^{(s)}_{22} (u,v) + G^{(s)}_{33} (u,v) \r ].
\end{align}
As before, it is convenient to introduce
\begin{equation}
G (u,v) = u^{\f{\D_1+\D_4}{4}}g (u,v),
\end{equation}
leading to the following crossing equations. 

For scalar probes
\begin{equation}
g^{(u)}\l ( \f{1}{u}, \f{v}{u} \r ) = g^{(s)}(u,v),
\end{equation}
for scalar-vector probes
\begin{align}
g^{(u)}_2 \l ( \f{1}{u}, \f{v}{u} \r ) & = - \l [ g^{(s)}_2 (u,v) + \f{1+u-v}{\sqrt{u}} g^{(u)}_3 (u,v)\r ], \\
g^{(u)}_3 \l ( \f{1}{u}, \f{v}{u} \r ) & = g^{(s)}_1 (u,v),
\end{align}
and for vector probes
\begin{align}
g^{(u)}_\d \l ( \f{1}{u}, \f{v}{u} \r ) & = g^{(s)}_\d (u,v), \\
g^{(u)}_{23} \l ( \f{1}{u}, \f{v}{u} \r ) & = g^{(s)}_{23} (u,v), \\
g^{(u)}_{22} \l ( \f{1}{u}, \f{v}{u} \r ) & = - \l [ g^{(s)}_{22} (u,v) + \f{1+u-v}{\sqrt{u}} g^{(s)}_{23} (u,v) + 2 g^{(s)}_{\d} (u,v) \r ], \\
g^{(u)}_{33} \l ( \f{1}{u}, \f{v}{u} \r ) & = - \l [ g^{(s)}_{33} (u,v) + \f{1+u-v}{\sqrt{u}} g^{(s)}_{23} (u,v) + 2 g^{(s)}_{\d} (u,v) \r ], \\
g^{(u)}_{32} \l ( \f{1}{u}, \f{v}{u} \r ) & = g^{(s)}_{32} (u,v) + \f{(1+u-v)^2}{u} g^{(s)}_{23} (u,v) \nn \\
& \quad \quad + \f{1+u-v}{\sqrt{u}} \l [ 2 g^{(s)}_{\d} (u,v) + g^{(s)}_{22} (u,v) + g^{(s)}_{33} (u,v) \r ].
\end{align}
Substituting,
\begin{equation}
\cos \t = \f{1+u-v}{2\sqrt{u}},
\end{equation}
we obtain for scalars
\begin{equation}
g^{(u)}\l ( \f{1}{u}, \f{v}{u} \r ) = g^{(s)}(u,v),
\end{equation}
scalar-vector
\begin{align}
g^{(u)}_2 \l ( \f{1}{u}, \f{v}{u} \r ) & = - \l [ g^{(s)}_2 (u,v) + 2 g^{(u)}_3 (u,v) \cos \t \r ], \\
g^{(u)}_3 \l ( \f{1}{u}, \f{v}{u} \r ) & = g^{(s)}_3 (u,v),
\end{align}
and vectors
\begin{align}
g^{(u)}_\d \l ( \f{1}{u}, \f{v}{u} \r ) & = g^{(s)}_\d (u,v), \\
g^{(u)}_{23} \l ( \f{1}{u}, \f{v}{u} \r ) & = g^{(s)}_{23} (u,v), \\
g^{(u)}_{22} \l ( \f{1}{u}, \f{v}{u} \r ) & = - \l [ g^{(s)}_{22} (u,v) + 2 g^{(s)}_{23} (u,v) \cos \t  + 2 g^{(s)}_{\d} (u,v) \r ], \\
g^{(u)}_{33} \l ( \f{1}{u}, \f{v}{u} \r ) & = - \l [ g^{(s)}_{33} (u,v) + 2 g^{(s)}_{23} (u,v) \cos \t + 2 g^{(s)}_{\d} (u,v) \r ], \\
g^{(u)}_{32} \l ( \f{1}{u}, \f{v}{u} \r ) & = g^{(s)}_{32} (u,v) + 4 g^{(s)}_{23} (u,v) \cos^2 \t \nn \\
& \quad \quad + 2 \cos \t \l [ 2 g^{(s)}_{\d} (u,v) + g^{(s)}_{22} (u,v) + g^{(s)}_{33} (u,v) \r ].
\end{align}

\subsection{Large-charge limit and bootstrap functions}

We now specialize to the heavy-light setup relevant for the large-charge bootstrap. We take
\begin{align}
\D_1 = \D_4 = \D_Q, ~~ \D = \D_Q +\omega_\ell, ~~ \D_{\phi_2} = \D_{\phi_3} = \d_S, ~~ \D_V = \d_V.
\end{align}
Using the $s$-channel expansions
\eqref{eq:ExplicitExpansionScalars}-\eqref{eq:ExplicitExpansionVectors}, we define the functions
$g^Q$ and $g^Q_I$ (with $I \in\{2,3,22,23,32,33,\delta\}$) for $\tau<0$ as follows.

For scalars
\begin{align}
g^Q(\tau, \t) & = | \lambda_{S,Q}|^2 \l [1 +e^{\tau} \f{\d_S^2}{2 \D_Q} \cos\t \r ] 
+ \sum_\ell \f{\ell!}{2^{\ell} \l(\f{d}{2} -1 \r )_\ell} |\lambda_{S,\ell}|^2 \, e^{\omega_\ell \tau} \, C^{(d/2-1)}_\ell(\cos\t)
\end{align}
for scalar-vector
\begin{align}
\label{eq:g3-SJ}
g^Q_2(\tau,\t) & = \lambda_{V,Q}^* \lambda_{S,Q}
\l (1 + e^{\tau} \f{\d_S (\d_V-1)}{2 \D_Q} \cos \t\r ) \\
& \quad \quad 
+ \l ( d-2 \r ) \sum_\ell \f{\ell!}{2^{\ell} \l(\f{d}{2} -1 \r )_\ell} \lambda_{S,\ell} \, e^{\omega_\ell \tau } 
\l [ \f{\lambda^+_{\ell}}{2 \ell+ d - 2} C_{\ell}^{(d/2)}(\cos\t) - \lambda^-_{\ell} C_{\ell-2}^{(d/2)}(\cos\t) \r ], \nn \\
\label{eq:g3-Scalars}
g^Q_3(\tau,\t) & = \lambda_{V,Q}^* \lambda_{S,Q} \, \f{\d_S}{2 \D_Q} \, e^{\tau} - \l ( d-2 \r ) \sum_\ell \f{\ell!}{2^{\ell} \l(\f{d}{2} -1 \r )_\ell} \lambda_{S,\ell} \, 
e^{\omega_\ell \tau } \l [ \f{\lambda^+_{\ell}}{2 \ell+ d - 2} - \lambda^-_{\ell} \r ] C_{\ell-1}^{(d/2)}(\cos\t).
\end{align}
and for vectors
\begin{align}
g^Q_{23}(\tau,\t) & = d(d-2)\sum_\ell \f{\ell!}{2^{\ell} \l(\f{d}{2} -1 \r )_\ell} \, e^{\omega_\ell \tau} \, c_{23,\ell}, \\
g^Q_{\d}(\tau,\t)& = \f{|\lambda_{V,Q}|^2}{2\D_Q} e^{\tau} + d (d-2) \sum_\ell \f{\ell!}{2^{\ell} \l(\f{d}{2} -1 \r )_\ell} \, e^{\omega_\ell \tau} \, c_{\d,\ell}, \\
g^Q_{22}(\tau,\t)& = \f{ |\lambda_{V,Q}|^2}{2\D_Q} (\d_V-1) e^{\tau} + d (d-2) \sum_\ell \f{\ell!}{2^{\ell} \l(\f{d}{2} -1 \r )_\ell} \, e^{\omega_\ell \tau} \, c_{22,\ell}, \\
g^Q_{33}(\tau,\t)& = -\f{ |\lambda_{V,Q}|^2}{2\D_Q} (\d_V+1) e^{\tau} + d (d-2) \sum_\ell \f{\ell!}{2^{\ell} \l(\f{d}{2} -1 \r )_\ell} \, e^{\omega_\ell \tau} \, c_{33,\ell}, \\
g^Q_{32}(\tau,\t)& =
-  |\lambda_{V,Q}|^2 \l [ 1 + \f{(\d_V^2-1)}{2\D_Q} \, e^{\tau} \cos \t \r ] 
+ d (d-2) \sum_\ell \f{\ell!}{2^{\ell} \l(\f{d}{2} -1 \r )_\ell} \, e^{\omega_\ell \tau} \, c_{32,\ell}.
\end{align}

In this heavy-light configuration the crossing equations simplify in cylinder coordinates. For scalars the crossing relation reduces to
\begin{equation}
g^{-Q}(-\tau, \t) = g^Q(\tau, \t),
\end{equation}
for scalar-vector we find
\begin{align}
g^{-Q}_2 (-\tau,\t) & = - \Big [ g^Q_2 (\tau,\t) + 2 g^Q_3 (\tau,\t) \cos\t \Big ], \\
g^{-Q}_3 (-\tau,\t) & = g^Q_3 (\tau,\t), 
\end{align}
and for vectors
\begin{align}
g^{-Q}_\d ( -\tau,\t) & = g^Q_\d (\tau,\t), \qquad g_{23} \l ( -\tau,\t \r ) = g^Q_{23} (\tau,\t), \\
g^{-Q}_{22} ( -\tau,\t ) & = - \Big [g^Q_{22} (\tau,\t) + 2 g^Q_{23} (\tau,\t) \cos\t  + 2 g^Q_{\d} (\tau,\t) \Big ], \\
g^{-Q}_{33} ( -\tau,\t ) & = - \Big [ g^Q_{33} (\tau,\t) + 2 g^Q_{23} (\tau,\t) \cos\t + 2 g^Q_{\d} (\tau,\t) \Big ], \\
g^{-Q}_{32}( -\tau,\t ) & = g^Q_{32} (\tau,\t) + 2 \Big [  2 g^Q_{\d} (\tau,\t) + g^Q_{22} (\tau,\t) + g^Q_{33} (\tau,\t) \Big ] \cos\t + 4g^Q_{23} (\tau,\t) \cos^2\t .
\end{align}

\subsection{Bootstrap parametrization and macroscopic-limit basis}

At leading order in $Q^{-1}$ the crossing equations impose the following constraint
\begin{align}
|\lambda_{S,-Q}|^2 = |\lambda_{S,Q}|^2, \quad |\lambda_{V,-Q}|^2 = |\lambda_{V,Q}|^2, \quad \lambda_{V,-Q}^*\lambda_{S,-Q} = -\lambda_{V,Q}^*\lambda_{S,Q}.
\end{align}
To isolate the subleading constraints we factor out the leading pieces and define, for scalars
\begin{equation}
g^Q (\tau,\t) = | \lambda_{S,Q}|^2 \left[1 + \frac{\d_S^2}{2 \D_Q}\, f(\tau,\t) \right],
\end{equation}
for scalar-vector
\begin{align}
\label{eq:SVgf}
g^Q_2(\tau,\t) & = \lambda_{V,Q}^* \lambda_{S,Q} 
\l (1 + \f{\d_S \d_V}{2 \D_Q} f_{2} (\tau,\t) \r ) \\
g^Q_3(\tau,\t) & = \lambda_{V,Q}^* \lambda_{S,Q} \, \f{\d_S \d_V}{2 \D_Q} f_{3} (\tau,\t) \nn
\end{align}
and vectors
\begin{align}
\label{eq:VVgf}
g_{23}(\tau,\t) & = \f{|\lambda_{V,Q}|^2 \d_V^2}{2\D_Q} f_{23}(\tau,\t), ~~ g_{\d}(\tau,\t) = \f{|\lambda_{V,\D_Q}|^2 \d_V^2}{2\D_Q} f_{\d}(\tau,\t) \\
g_{22}(\tau,\t)& = \f{|\lambda_{V,Q}|^2 \d_V^2}{2\D_Q} f_{22}(\tau,\t), ~~ g_{33}(\tau,\t) = \f{|\lambda_{V,\D_Q}|^2 \d_V^2}{2\D_Q} f_{33}(\tau,\t), \nn \\
g_{32}(\tau,\t)& =
- |\lambda_{V,Q}|^2 \l [ 1 + \f{ \d_V^2}{2\D_Q} f_{32}(\tau,\t)\r ]. \nn
\end{align}
Using that all subleading fusion coefficients scale as $\D_Q^{-1/2}$, it is convenient to redefine
\begin{equation}
\lambda_{S,\ell} = \f{\lambda_{S,Q} \d_S}{\sqrt{2 \D_Q}} \, \sqrt{\f{2^{\ell} \l(\f{d}{2} -1 \r )_\ell}{\ell!}}  \, \m_{S,\ell}
\end{equation}
and
\begin{equation}
\lambda^{\pm}_\ell = \f{ \lambda^*_{V,Q} \d_V}{\sqrt{2\D_Q}} \, \sqrt{\f{2^{\ell} \l(\f{d}{2} -1 \r )_\ell}{\ell!}} \, \n^{\pm}_\ell, 
\end{equation}
which leads to the following parametrization at order $1/\D_Q$.

For scalars
\begin{align}
\label{eq:f-Scalars}
f(\tau, \t) & = e^{\tau} \cos\t 
+ \sum_\ell |\m_{S,\ell}|^2 \, e^{\omega_\ell \tau} \, C^{(d/2-1)}_\ell(\t)
\end{align}
for scalar-vector
\begin{align}
f_2(\tau,\t) & = 
(1-\d_V^{-1}) e^{\tau} \cos \t + \l ( d-2 \r ) \sum_\ell \m_{S,\ell} \, e^{\omega_\ell \tau } 
\l [ \f{\n^+_{\ell}}{2 \ell+ d - 2} C_{\ell}^{(d/2)}(\cos \t) - \n^-_{\ell} C_{\ell-2}^{(d/2)}(\cos \t) \r ], \nn \\
f_3(\tau,x) & = \d_V^{-1}e^{\tau} -  \l ( d-2 \r ) \sum_\ell \m_{S,\ell} \, 
e^{\omega_\ell \tau } \l [ \f{\n^+_{\ell}}{2 \ell+ d - 2} - \n^-_{\ell} \r ] C_{\ell-1}^{(d/2)}(\cos \t),
\end{align}
and for vectors
\begin{align}
f_{23}(\tau,\t) & = d(d-2)\sum_\ell e^{\omega_\ell \tau} \, c_{23,\ell}(\n^+_\ell,\n^-_\ell), \\
f_{\d}(\tau,\t)& = \d_V^{-2}e^{\tau} + d (d-2) \sum_\ell e^{\omega_\ell \tau} \, c_{\d,\ell}(\n^+_\ell,\n^-_\ell), \\
f_{22}(\tau,\t)& = \d_V^{-2}(\d_V-1) e^{\tau} + d (d-2) \sum_\ell e^{\omega_\ell \tau} \, c_{22,\ell}(\n^+_\ell,\n^-_\ell), \\
f_{33}(\tau,\t)& = - \d_V^{-2}(\d_V+1) e^{\tau} + d (d-2) \sum_\ell e^{\omega_\ell \tau} \, c_{33,\ell}(\n^+_\ell,\n^-_\ell), \\
f_{32}(\tau,\t)& =
- (1-\d_V^{-2}) \, e^{\tau} \cos \t 
+ d (d-2) \sum_\ell e^{\omega_\ell \tau} \, c_{32,\ell}(\n^+_\ell,\n^-_\ell).
\end{align}

Crossing equations for these functions are as follows, for scalars
\begin{equation}
f(-\tau, x) =  f(\tau, x),
\end{equation}
for scalar-vector\footnote{We assume that fusion coefficients corresponding to the subleading primaries have the $Q$-dependence as the leading order operator, which also implies that they have the same $C$-parity.}
\begin{align}
f_2 (-\tau,\t) & = f_2 (\tau,\t) + 2 x f_3 (\tau,\t), \\
f_3 (-\tau,\t) & = - f_3 (\tau,\t), 
\end{align}
and for vectors
\begin{align}
f_\d ( -\tau,\t ) & = f_\d (\tau,\t), ~~ f_{23} \l ( -\tau,\t \r ) = f_{23} (\tau,\t), \\
f_{22} ( -\tau,\t ) & = - \Big [f_{22} (\tau,\t) + 2 f_{23} (\tau,x) \cos \t + 2 f_{\d} (\tau,\t) \Big ], \\
f_{33} ( -\tau,\t) & = - \Big [ f_{33} (\tau,\t) + 2 f_{23} (\tau,\t) \cos \t  + 2 f_{\d} (\tau,\t) \Big ], \\
f_{32}( -\tau,\t ) & = f_{32} (\tau,\t) + 2 \Big [  2 f_{\d} (\tau,\t) + f_{22} (\tau,\t) + f_{33} (\tau,\t) \Big ]\cos\t + 4 f_{23} (\tau,\t) \cos^2 \t.
\end{align}

Finally, to make the macroscopic limit manifest and to parallel the scalar analysis, we define the following functions.

For scalars
\begin{align}
h (\tau, \t) & = f(\tau, \t),
\end{align}
for scalar-vector
\begin{align}
\label{eq:SVfh}
h_{2} (\tau, \t) & = f_2 (\tau,\t) + f_3(\tau,\t) \cos \t, \\
h_{3} (\tau, \t) & = f_3(\tau,\t), \nn
\end{align}
and for vectors
\begin{align}
\label{eq:VVfh}
h_{23} (\tau, \t) & = f_{23}(\tau,\t), \quad h_{\d} (\tau, \t) = f_{\d}(\tau,\t) \\
h_{22} (\tau, \t) & = f_{22}(\tau,\t) + f_{23}(\tau,\t) \cos\t + f_{\d}(\tau,\t) \\
-h_{33} (\tau, \t) & = f_{33}(\tau,\t) + f_{23}(\tau,\t) \cos\t + f_{\d}(\tau,\t) \\
-h_{32} (\tau, \t) & = f_{32}(\tau,x) + \Big [ f_{22}(\tau,\t) + f_{33}(\tau,\t) + f_{\d}(\tau,\t) \Big ] \cos \t + f_{23}(\tau,\t) \cos^2\t,
\end{align}
with $h(\tau, \t)$, $h_{2}(\tau, \t)$, $h_{23}(\tau, \t)$, $h_{\d}(\tau, \t)$, $h_{32}(\tau, \t)$ being even and $h_{3}(\tau, \t)$, $h_{22}(\tau, \t)$, $h_{33}(\tau, \t)$, being odd functions in $\tau$,
and having the following representation in terms of the Gegenbauer polynomials.

For scalars
\begin{align}
\label{eq:hSSmuExpansion}
h (\tau, x) & = e^{\tau} \cos\t 
+ \sum_\ell |\m_{S,\ell}|^2 \, e^{\omega_\ell \tau} \, C^{(d/2-1)}_\ell(\t),
\end{align}
for scalar-vector
\begin{align}
\label{eq:hSVnuExpansion}
h_{2} (\tau, \t) & = e^{\tau} \cos \t 
+ \sum_\ell \m_{S,\ell} \l (\f{\ell+d-2}{2 \ell+ d - 2}\n^+_{\ell} + \ell \n^-_{\ell} \r ) e^{\omega_\ell \tau} \, C_{\ell}^{(d/2-1)}(\cos \t), \\
h_{3} (\tau, \t) & = \d_V^{-1}e^{\tau}
- (d-2) \sum_\ell \m_{S,\ell} \l ( \f{\n^+_{\ell} }{2 \ell+ d - 2} - \n^-_{\ell} \r ) e^{\omega_\ell \tau} \, C_{\ell-1}^{(d/2)}(\cos \t),
\end{align}
and for vectors
\begin{align}
\label{eq:hVVnuExpansion}
h_{23} (\tau, \t) & = d(d-2) \sum_\ell \l ( \f{\n^+_{\ell} }{2 \ell+ d - 2} - \n^-_{\ell} \r ) 
\l ( \f{\bar \n^+_{\ell} }{2 \ell+ d - 2} - \bar \n^-_{\ell} \r ) e^{\omega_\ell \tau} \, C_{\ell-2}^{(d/2+1)}(\cos \t), \\
h_{\d} (\tau, \t) & = \d_V^{-2}e^{\tau} + (d-2) \sum_\ell \l ( \f{\n^+_{\ell} }{2 \ell+ d - 2} - \n^-_{\ell} \r ) 
\l ( \f{\bar \n^+_{\ell} }{2 \ell+ d - 2} - \bar \n^-_{\ell} \r ) e^{\omega_\ell \tau} \, C_{\ell-1}^{(d/2)}(\cos \t), \nn \\
h_{22} (\tau, \t) & = \d_V^{-1}e^{\tau} - (d-2) \sum_\ell 
\l (\f{\ell+d-2}{2 \ell+ d - 2}\n^+_{\ell} + \ell \n^-_{\ell} \r ) \l ( \f{\bar \n^+_{\ell} }{2 \ell+ d - 2} - \bar \n^-_{\ell} \r ) 
e^{\omega_\ell \tau} \, C_{\ell-1}^{(d/2)}(\cos \t), \nn \\
h_{33} (\tau, \t) & = \d_V^{-1}e^{\tau} + (d-2) \sum_\ell 
\l ( \f{\n^+_{\ell} }{2 \ell+ d - 2} - \n^-_{\ell} \r ) \l (\f{\ell+d-2}{2 \ell+ d - 2}\bar \n^+_{\ell} + \ell \bar \n^-_{\ell} \r ) 
e^{\omega_\ell \tau} \, C_{\ell-1}^{(d/2)}(\cos \t), \nn \\
h_{32} (\tau, \t) & = e^{\tau} \cos \t - \sum_\ell 
\l (\f{\ell+d-2}{2 \ell+ d - 2}\n^+_{\ell} + \ell \n^-_{\ell} \r ) \l (\f{\ell+d-2}{2 \ell+ d - 2}\bar \n^+_{\ell} + \ell \bar \n^-_{\ell} \r )
e^{\omega_\ell \tau} \, C_{\ell}^{(d/2-1)}(\cos \t). \nn
\end{align}
Equivalently,
\begin{align}
\label{eq:hVVnuExpansionLambda}
h_{23} (\tau, \t) & = d(d-2) \sum_\ell 
\l | \f{\n^+_{\ell} }{2 \ell+ d - 2} - \n^-_{\ell} \r |^2 
e^{\omega_\ell \tau} \, C_{\ell-2}^{(d/2+1)}(\cos \t), \\
h_{\d} (\tau, \t) & = \d_V^{-2}e^{\tau} + (d-2) \sum_\ell 
\l | \f{\n^+_{\ell} }{2 \ell+ d - 2} - \n^-_{\ell} \r |^2 
e^{\omega_\ell \tau} \, C_{\ell-1}^{(d/2)}(\cos \t), \nn \\
h_{22} (\tau, \t) & = \d_V^{-1}e^{\tau} - (d-2) \sum_\ell 
\l [ \f{\ell+d-2}{2 \ell+ d - 2}\n^+_{\ell} + \ell \n^-_{\ell} \r ] \l [ \f{(\n^+_{\ell})^* }{2 \ell+ d - 2} - (\n^-_{\ell})^* \r ] 
e^{\omega_\ell \tau} \, C_{\ell-1}^{(d/2)}(\cos \t), \nn \\
h_{33} (\tau, \t) & = \d_V^{-1}e^{\tau} - (d-2) \sum_\ell 
\l [ \f{\n^+_{\ell} }{2 \ell+ d - 2} - \n^-_{\ell} \r ] \l [ \f{\ell+d-2}{2 \ell+ d - 2} (\n^+_{\ell})^* + \ell (\n^-_{\ell})^* \r ] 
e^{\omega_\ell \tau} \, C_{\ell-1}^{(d/2)}(\cos \t), \nn \\
h_{32} (\tau, \t) & = e^{\tau} \cos \t + \sum_\ell 
\l |\f{\ell+d-2}{2 \ell+ d - 2}\n^+_{\ell} + \ell \n^-_{\ell} \r |^2
e^{\omega_\ell \tau} \, C_{\ell}^{(d/2-1)}(\cos \t). \nn
\end{align}
For the conserved current using \eqref{eq:LambdaRatio} these equations reduce to \eqref{eq:hVVConserved}.

%

\section{Four-point functions in EFT}

Even though the results of this section were previously obtained in~\cite{Loic:2016, Dondi:2022wli} we present them here to facilitate referencing.

\subsection{Two scalars}

Here we consider the four point function corresponding to the insertion of two neutral scalars
\begin{equation}
\la Q | \hat {\mc O}_3 (\tau_3, \vec n_3) \hat {\mc O}_2 (\tau_2, \vec n_2) | Q \ra. 
\end{equation}
With the operator $\hat {\mc O}$ at leading order according to \eqref{eq:EFT-operator-matching} given by
\begin{equation}
\hat {\mc O} = C_\mc O \l [ - \l ( \p \c \r )^2 \r ]^{\d/2}.
\end{equation}
Expanding up to linear order the operators $\hat {\mc O}_{2,3}$, we obtain
\begin{align}
\la Q | \hat {\mc O}_3 (\tau_3, \vec n_3) \hat {\mc O}_2 (\tau_2, \vec n_2) | Q \ra & = 
C_2 C_3 \m^{\d_2+\d_3} \l [ 1-\f{\d_2 \d_3}{c_d d (d-1)\m^{d}} \la \dot {\pi}_3 \dot {\pi}_2 \ra \r ].
\end{align}
Using that the propagator on the cylinder is given by\footnote{Here we use the generalized form of the addition theorem for spherical harmonics~\cite{Frye:2012jj}
\begin{equation}
\sum_{\vec m}Y_{\ell,\vec m}(\vec n_2)Y^*_{\ell,\vec m}(\vec n_1) = \f{2\ell+d-2}{(d-2) \Omega_{d-1}} C^{(d/2-1)}_\ell(\vec n_1 \vec n_2)
\end{equation}}
\begin{equation}
\la \pi_2 \pi_1 \ra = \sum_{\ell} \f{e^{-\omega_\ell(\tau_2-\tau_1)/R}}{2\omega_\ell} \f{2\ell+d-2}{(d-2) \Omega_{d-1} R^{d-2}} 
C^{(d/2-1)}_\ell(\vec n_1 \vec n_2),
\end{equation}
we can compute
\begin{align}
\la \dot {\pi}_3 \dot {\pi}_2 \ra & = -\f{1}{2(d-2)\Omega_{d-1}R^{d}} \sum_\ell (2\ell+d-2) \omega_\ell \, e^{-\omega_\ell(\tau_3-\tau_2)/R} C^{(d/2-1)}_\ell(\vec n_2 \vec n_3).
\end{align}
As a result, using \eqref{eq:QmuR-Relation} we obtain
\begin{align}
&\la Q | \hat {\mc O}_3 (\tau_3, \vec n_3) \hat {\mc O}_2 (\tau_2, \vec n_2) | Q \ra 
= \la Q | \tilde {\mc O}_3 (\tau_3, \vec n_3) \tilde {\mc O}_2 (\tau_2, \vec n_2) | Q \ra \\
& = C_2 C_3 \m^{\d_2+\d_3} \l [ 1+\f{\d_2 \d_3}{2 c_d d (d-1) (d-2) \Omega_{d-1} (\m R)^{d} } \sum_{\ell=1}^\infty (2\ell+d-2) \omega_\ell \, e^{-\omega_\ell(\tau_3-\tau_2)/R} C^{(d/2-1)}_\ell(\vec n_2 \vec n_3) \r ] \nn \\
& = C_2 C_3 \m^{\d_2+\d_3} \l [ 1+\f{\d_2 \d_3}{2 d (d-2) \D_Q}  \sum_{\ell=1}^\infty (2\ell+d-2) \omega_\ell \, e^{-\omega_\ell(\tau_3-\tau_2)/R} C^{(d/2-1)}_\ell(\vec n_2 \vec n_3) \r ] \nn \\
& = C_2 C_3 \m^{\d_2+\d_3} \l [ 1 + \f{\d_2 \d_3}{2 \D_Q} |z| \f{C^{(d/2-1)}_\ell(\vec n_2 \vec n_3)}{(d-2)} \r ] \\ 
& \quad \quad + C_2 C_3 \m^{\d_2+\d_3} \f{\d_2 \d_3}{2 d (d-2) \D_Q}  \sum_{\ell=2}^\infty (2\ell+d-2) \omega_\ell \, e^{-\omega_\ell(\tau_3-\tau_2)/R} C^{(d/2-1)}_\ell(\vec n_2 \vec n_3), \nn
\end{align}
leading to
\begin{align}
f^{EFT}(\tau,x) & = e^\tau \f{C^{(d/2-1)}_\ell(\vec n_2 \vec n_3)}{(d-2)} + \f{1}{d (d-2)}  \sum_{\ell=2}^\infty (2\ell+d-2) \omega_\ell \, e^{-\omega_\ell(\tau_3-\tau_2)/R} C^{(d/2-1)}_\ell(\vec n_2 \vec n_3),
\end{align}
equivalently
\begin{align}
\label{eq:hEFT}
h^{EFT}(\tau,x) & = e^\tau \cos \t + \f{1}{d (d-2)}  \sum_{\ell=2}^\infty (2\ell+d-2) \omega_\ell \, e^{-\omega_\ell(\tau_3-\tau_2)/R} C^{(d/2-1)}_\ell(\cos \t).
\end{align}

\subsection{Scalar-current}

First, we consider the four point function of the form \eqref{eq:4ptFunctionSSVS-generic} involving a neutral scalar operator. On the cylinder, this four point function can be computed by considering the following matrix element
\begin{equation}
\la Q | \hat {\mc O} (\tau_3, \vec n_3) \hat J^\m (\tau_2, \vec n_2) | Q \ra. 
\end{equation}
With the operator $\hat {\mc O}$ at leading order according to \eqref{eq:EFT-operator-matching} given by
\begin{equation}
\hat {\mc O} = C_\mc O \l [ - \l ( \p \c \r )^2 \r ]^{\d/2}.
\end{equation}
Expanding up to linear order both the current and the operator $\hat {\mc O}$, we obtain
\begin{align}
\la Q | \hat {\mc O} (\tau_3, \vec n_3) \hat J^\m (\tau_2, \vec n_2) | Q \ra 
& = i C_0 \m^\d \f{Q}{\Omega_{d-1}R^{d-1}} 
\l \{ \d^\m _\tau - \f{\d}{c_dd(d-1)\m^d} \Big [ \la \dot \pi_3 \p^\m \pi_2 \ra + (d-2) \d^\m _\tau  \la \dot \pi_3 \dot \pi _2\ra \Big ] \r \} \nn \\
& =
i C_0 \m^\d \f{Q}{\Omega_{d-1}R^{d-1}} \d^\m _\tau - i C_0 \m^\d \f{Q}{\Omega_{d-1}R^{d-1}} \f{\d}{c_dd(d-1)\m^d} \hat I_4^\m,
\end{align}
with
\begin{equation}
\hat I_4^\m = \la \dot \pi_3 \p^\m \pi_2 \ra + (d-2) \d^\m _\tau \la \dot \pi_3 \dot \pi _2\ra.
\end{equation}
Using that the propagator on the cylinder is given by
\begin{equation}
\la \pi_2 \pi_1 \ra = \sum_{\ell} \f{e^{-\omega_\ell(\tau_2-\tau_1)/R}}{2\omega_\ell} \f{2\ell+d-2}{(d-2) \Omega_{d-1} R^{d-2}} 
C^{(d/2-1)}_\ell(\vec n_1 \vec n_2),
\end{equation}
we can compute
\begin{align}
\hat I_4^\tau & = -\f{d-1}{2(d-2)\Omega_{d-1}R^{d}} \sum_\ell (2\ell+d-2) \omega_\ell \, e^{-\omega_\ell(\tau_3-\tau_2)/R} C^{(d/2-1)}_\ell(\vec n_2 \vec n_3), \\
\hat I_4^i & = -\f{1}{2(d-2)\Omega_{d-1}R^{d+1}} \sum_\ell (2\ell+d-2) \, e^{-\omega_\ell(\tau_3-\tau_2)/R} G^{ij}(\vec n_2) \p_j^{(2)} C^{(d/2-1)}_\ell(\vec n_2 \vec n_3).
\end{align}
As a result,
\begin{align}
\tilde I_4^a & = \f{R}{x_2} \f{\p x_2^a}{\p x_2^\m} \hat I_4^\m = n_2^a \hat I_4^\tau + R \f{\p n_2^a}{\p \t_2^i} \hat I_4^i \\
& = -\f{1}{2(d-2)\Omega_{d-1}R^d} \sum_\ell (2\ell+d-2) \, e^{-\omega_\ell(\tau_3-\tau_2)/R} 
\l [ (d-1) \omega_\ell n_2^a + (g^{ab} - n_2^a n_2^b) \f{\p}{\p n^b_2} \r ] C^{(d/2-1)}_\ell(\vec n_2 \vec n_3) \nn \\
& = -\f{1}{2(d-2)\Omega_{d-1}R^d} \sum_\ell (2\ell+d-2) \, e^{-\omega_\ell(\tau_3-\tau_3)/R} 
\l [ (d-1) \omega_\ell n_2^a C^{(d/2-1)}_\ell + (n_3^{a} - n_2^a \vec n_2 \vec n_3) C^{(d/2-1)'}_\ell \r ]  \nn
\end{align}
Using the recursion relations \eqref{eq:GegenbauerRecursion1}-\eqref{eq:GegenbauerRecursion3} for the Gegenbauer polynomials, we obtain
\begin{align}
-2 \tilde I_4^a \Omega_{d-1} R^d & = n_2^a\sum_\ell e^{-\omega_\ell(\tau_3-\tau_2)/R} 
\bigg \{ C^{(d/2)}_\ell(\vec n_2 \vec n_3) \Big [ (d-1) \omega_\ell -\ell \Big ] - C^{(d/2)}_{\ell-2}(\vec n_2 \vec n_3) \Big[ (d-1) \omega_\ell +\ell+d-2 \Big ] \bigg \} \nn \\
& \quad \quad + n_3^a\sum_\ell (2\ell+d-2) \, e^{-\omega_\ell(\tau_3-\tau_2)/R} C^{(d/2)}_{\ell-1}(\vec n_2 \vec n_3),
\end{align}
leading to
\begin{align}
\la Q | \tilde {\mc O} (\tau_3, \vec n_3) \tilde J^a (\tau_2, \vec n_2) | Q \ra
& = i C_0 \m^\d \f{Q}{\Omega_{d-1}R^{d-1}} \l \{ n_2^a + e^{-(\tau_3-\tau_2)/R} \l [ \f{\d (d-2)}{2\D_Q} \f{C^{(d/2)}_1(\vec n_3 \vec n_2)}{d}n_2^a + \f{\d}{2\D_Q} n_3^a \r] \r \} \nn \\
& + i C_0 \m^\d \f{Q}{\Omega_{d-1}R^{d-1}} \f{\d}{2\D_Q d} \Bigg \{ n_2^a\sum_{\ell=2}^\infty e^{-\omega_\ell(\tau_3-\tau_2)/R} 
\bigg \{ C^{(d/2)}_\ell(\vec n_3 \vec n_2) \Big [ (d-1) \omega_\ell -\ell \Big ] \nn \\ 
& \quad \quad \quad \quad - C^{(d/2)}_{\ell-2}(\vec n_2 \vec n_3) \Big[ (d-1) \omega_\ell +\ell+d-2 \Big ] \bigg \} \nn \\
& \quad \quad + n_3^a\sum_{\ell=2}^\infty (2\ell+d-2) \, e^{-\omega_\ell(\tau_3-\tau_2)/R} C^{(d/2)}_{\ell-1}(\vec n_2 \vec n_3) \Bigg \},
\end{align}

As a result, we have the following functions
\begin{align}
f_2^{EFT}(\tau,\t)
& = \f{d-2}{d-1} \, e^{\tau} \, \cos \t \\
& \quad \quad + \f{1}{d(d-1)} \sum_{\ell=2}^\infty e^{\omega_\ell \tau} 
\bigg \{ C^{(d/2)}_\ell(\cos \t) \Big [ (d-1) \omega_\ell -\ell \Big ] - C^{(d/2)}_{\ell-2}(\cos \t) \Big[ (d-1) \omega_\ell +\ell+d-2 \Big ] \bigg \}, \nn \\
f_3^{EFT}(\tau,\t)
& = \f{e^{\tau}}{d-1} + \f{1}{d(d-1)} \sum_{\ell=2}^\infty (2\ell+d-2) \, e^{\omega_\ell \tau} C^{(d/2)}_{\ell-1}(\cos \t),
\end{align}
and
\begin{align}
\label{eq:h2EFT}
h_2^{EFT}(\tau,\t) & = e^{\tau} \cos\t +\f{1}{d(d-2)} \sum_{\ell=2}^\infty 
(2\ell+d-2)\omega_\ell \, e^{\omega_\ell \tau} C^{(d/2-1)}_{\ell}(\cos\t), \\
\label{eq:h3EFT}
h_3^{EFT}(\tau,\t) & = \f{e^{\tau}}{d-1} + \f{1}{d(d-1)} \sum_{\ell=2}^\infty 
(2\ell+d-2) \, e^{\omega_\ell \tau} C^{(d/2-1)}_{\ell}(\cos\t).
\end{align}

\subsection{Two currents}

Using the same expression for the current \eqref{eq:CurrentEFT}, we get
\begin{align}
\la Q | \tilde J^a (x_3) \tilde J^b (x_2) | Q \ra & = - \l (\f{Q}{\Omega_{d-1}}\r )^2 n_3^a n_2^b + \l (\f{Q}{\Omega_{d-1}}\r )^2 \f{\tilde I^{ab}}{c_d d (d-1)\m^d},
\end{align}
with
\begin{align}
\tilde I^{ab} & = \f{1}{2(d-2)\Omega_{d-1}} \sum_\ell \f{2\ell+d-2}{\omega_\ell} e^{-\omega_\ell(\tau_2-\tau_1)} \\
& =
\bigg [ - (d-1)^2 \omega_\ell^2 n_3^a n_2^b - (d-1)\omega_\ell \, n_3^a \l ( \d^{bd} - n_2^b n_2^d \r ) \f{\p}{\p n_2^d} + 
(d-1)\omega_\ell \, n_2^b \l ( \d^{ac} - n_3^a n_3^c \r ) \f{\p}{\p n_3^c} \nn \\
& \qquad \qquad+  \l ( \d^{ac} - n_3^a n_3^c \r ) \l ( \d^{bd} - n_2^b n_2^d \r ) \f{\p}{\p n_3^c} \f{\p}{\p n_2^d} \bigg ]
C_\ell^{(d/2-1)} (\vec n_2 \vec n_3).
\end{align}
Simplifying, it leads to
\begin{align}
\tilde I^{ab} & = \f{1}{2(d-2)\Omega_{d-1}} \sum_\ell \f{2\ell+d-2}{\omega_\ell} e^{-\omega_\ell(\tau_2-\tau_1)} \\
& \times
\bigg \{  
n_3^a n_2^b \Big [ - (d-1)^2 \omega_\ell^2 C_\ell^{(d/2-1)} (\vec n_2 \vec n_3) + (\vec n_2 \vec n_3) \l( C_\ell^{(d/2-1)} (\vec n_2 \vec n_3) \r)^{'} + (\vec n_2 \vec n_3)^2 \l( C_\ell^{(d/2-1)} (\vec n_2 \vec n_3) \r )^{''} \Big ] \nn \\
& \quad \quad - n_3^a n_3^b \bigg \{ \Big [ (d-1) \omega_\ell + 1 \Big ] \l ( C_\ell^{(d/2-1)} (\vec n_2 \vec n_3) \r )^{'} 
+ (\vec n_2 \vec n_3) \l ( C_\ell^{(d/2-1)} (\vec n_2 \vec n_3) \r )^{''} \bigg \} \nn \\
& \quad \quad + n_2^a n_2^b \bigg \{ \Big [ (d-1) \omega_\ell - 1 \Big ] \l ( C_\ell^{(d/2-1)} (\vec n_2 \vec n_3) \r )^{'} - (\vec n_2 \vec n_3) \l ( C_\ell^{(d/2-1)} (\vec n_2 \vec n_3) \r )^{''} \bigg \}   \nn \\
& \quad \quad + n_3^b n_2^a \l( C_\ell^{(d/2-1)} (\vec n_2 \vec n_3) \r )^{''} + \d^{ab} \l ( C_\ell^{(d/2-1)}(\vec n_2 \vec n_3) \r )^{'} \bigg \},
\end{align}
which upon using relations for Gegenbauer polynomials results in
\begin{align}
\la Q | \tilde J^a (x_3) \tilde J^b (x_2) | Q \ra & =  \l (\f{Q}{\Omega_{d-1}}\r )^2 \Bigg \{ - n_3^a n_2^b \l [1 + 
|z| \f{d(d-2)}{2\D_Q} \f{C_1^{(d/2+1)}(\vec n_2 \vec n_3)}{d+2} \r ] \nn \\
& \quad \quad \quad \quad \quad \quad \quad \quad + \f{|z|}{2\D_Q} \bigg [\d^{ab} - d n_3^{a} n_3^b + (d-2) n_2^{a} n_2^b \bigg ] \Bigg \} \nn \\
& \quad + \l (\f{Q}{\Omega_{d-1}}\r )^2 \f{1}{2\D_Q} \sum_{\ell=2} ^ \infty  \f{e^{-\omega_\ell(\tau_3-\tau_2)}}{\omega_\ell} \nn \\
& \quad \quad \times \bigg \{ \d^{ab} c_{\d,\ell}^{EFT} + n_3^{a} n_3^b c_{33,\ell}^{EFT} + n_3^{a} n_2^b c_{32,\ell}^{EFT} + n_2^{a} n_3^b c_{23,\ell}^{EFT} + n_2^{a} n_2^b c_{22,\ell}^{EFT} \bigg \} \nn
\end{align}
with
\begin{align}
c_{\d,\ell}^{EFT} & = \f{2\ell+d-2}{d}C_{\ell-1}^{(d/2)} (\vec n_2 \vec n_3), \\
c_{23,\ell}^{EFT} & = (2\ell+d -2) C_{\ell-2}^{(d/2+1)} (\vec n_2 \vec n_3), \\
c_{22,\ell}^{EFT} & = \big [ (d-1) \omega_\ell - \ell \big ] C_{\ell-1}^{(d/2+1)} (\vec n_2 \vec n_3) 
- \big [ (d-1) \omega_\ell + \ell + d - 2 \big] C_{\ell-3}^{(d/2+1)} (\vec n_2 \vec n_3), \\
c_{33,\ell}^{EFT} & = - \big [ (d-1) \omega_\ell + \ell \big ] C_{\ell-1}^{(d/2+1)} (\vec n_2 \vec n_3) 
+ \big [ (d-1) \omega_\ell - \ell - d + 2 \big] C_{\ell-3}^{(d/2+1)} (\vec n_2 \vec n_3), \\
c_{32,\ell}^{EFT} & = - \f{ (d-1)^2 \omega_\ell^2 - \ell^2 }{2\ell+d} \, C_{\ell}^{(d/2+1)} (\vec n_2 \vec n_3) \\
& \quad \quad \quad + 2 (2\ell+d-2) \f{ (d-1)^2 \omega_\ell^2 + \ell (\ell-2) + d (\ell-1) }{(2\ell+d)(2\ell+d-4)} \, C_{\ell-2}^{(d/2+1)} (\vec n_2 \vec n_3) \nn \\
& \quad \quad \quad \quad - \f{ \big [ (d-1) \omega_\ell + \ell + d - 2 \big ] \big [ (d-1) \omega_\ell - \ell - d + 2 \big ] }{2\ell+d-4} \, C_{\ell-4}^{(d/2+1)} (\vec n_2 \vec n_3). \nn
\end{align}
As a result,
\begin{align}
\label{eq:h23EFT}
h^{EFT}_{23} (\tau,\t) & = \f{1}{(d-1)^2} \sum_{\ell=2}^\infty \f{2\ell+d-2}{\omega_\ell} \, e^{\omega_\ell \tau} \, C_{\ell-2}^{(d/2+1)}(\cos\t), \\
\label{eq:hdeltaEFT}
h^{EFT}_{\d} (\tau,\t) & = \f{e^{\tau}}{(d-1)^2}+\f{1}{d(d-1)^2} \sum_{\ell=2}^\infty \f{2\ell+d-2}{\omega_\ell} \, e^{\omega_\ell \tau} \, C_{\ell-1}^{(d/2)}(\cos\t), \\
\label{eq:h22EFT}
h^{EFT}_{22} (\tau,\t) & =\f{e^{\tau}}{d-1} + \f{1}{d(d-1)} \sum_{\ell=2}^\infty (2\ell+d-2) \, e^{\omega_\ell \tau} \, C_{\ell-1}^{(d/2)}(\cos\t), \\
\label{eq:h33EFT}
h^{EFT}_{33} (\tau,\t) & =\f{e^{\tau}}{d-1} + \f{1}{d(d-1)} \sum_{\ell=2}^\infty (2\ell+d-2) \, e^{\omega_\ell \tau} \, C_{\ell-1}^{(d/2)}(\cos\t), \\
\label{eq:h32EFT}
h^{EFT}_{32} (\tau,\t) & = e^{\tau} \cos\t + 
\f{1}{d(d-2)} \sum_{\ell=2}^\infty (2\ell+d-2) \omega_\ell \, e^{\omega_\ell \tau} C_\ell^{(d/2-1)}(\cos \t).
\end{align}

\section{Laplacians in the embedding space}

For what follows we need to find eigenvalues for different Laplacian operators. To do that we reexpress them in terms of the embedding coordinates
\begin{equation}
x^a = r n^a (\t).
\end{equation}
The corresponding induced metric is
\begin{equation}
g_{ij} = e^a _i e^a_j = \p_i n^a \p_j n^a,
\end{equation}
and the connection
\begin{equation}
\G_{m,ik} = \p_m n^a \p_i \p _ k n^a = e^a _m \p_i e^a_j = e^a _m \p_j e^a_i
\end{equation}
Using the orthogonality
\begin{equation}
\p_i e^a _j n^a = - e^a _j \p_i n^a = - e^a _j e^a_i = -g_{ij},
\end{equation}
we derive
\begin{equation}
\nabla_i e_j^a = \p_i e_j^a - \G^k_{ij} e_k^a = \p_i e_j^a - g^{mk} e^b _m \p_i e^b_j e_k^a = \l ( \d^{ab} - P^{ab} \r) \p_i e_j^b,
\end{equation}
where we introduced the projector on the sphere
\begin{equation}
P^{ab} = g^{ij} e_i^a e^b _j = \p_a n^b = \p_b n^a = \d^{ab} - n^a n^b.
\end{equation}
The equation implies that $\nabla_i e_j^a$ is purely orthogonal, hence, since
\begin{equation}
\nabla_i e_j^a n^a = \p_i e_j^a n^a = -g_{ij},
\end{equation}
we get
\begin{equation}
\nabla_i e_j^a = - n^a g_{ij}.
\end{equation}
Using these expression we can show that for a scalar
\begin{equation}
P^{ab} \p_b \phi = g^{ij} e_i^a e^b_j  \p_b \phi = e_i^a \nabla^i \phi,
\end{equation}
equivalently
\begin{equation}
\label{eq:nablasigmaEmbedding}
\nabla_i \phi = e^a_i P^{ab} \p_b \phi.
\end{equation}
Similarly, introducing the vector field on a sphere as viewed from the ambient space
\begin{equation}
v^a = e^a_i v^i,
\end{equation}
we obtain
\begin{equation}
e^b_i \p_b v^a = \p_i v^a = \nabla_i e^a_j v^j + e^a_j \nabla_i v^j = - n^a g_{ij} v^j + e^a_j \nabla_i v^j,
\end{equation}
leading to
\begin{equation}
e^a_j \nabla_i v^j = e^c_i P^{ab} \p_c v^b,
\end{equation}
Equivalently,
\begin{equation}
\label{eq:nablavEmbedding}
\nabla_i v_j = e^a_j e^c_i P^{ab} \p_c v^b.
\end{equation}
Combining equations \eqref{eq:nablasigmaEmbedding} and \eqref{eq:nablavEmbedding}, we obtain\footnote{All derivatives are computed at $r=1$.}
\begin{align}
\label{eq:LaplacianSigma}
\nabla^2 \phi & = g^{ij}\nabla_i \nabla_j \phi = g^{ij} e^a_j e^c_i P^{ab} \p_c \l ( e^b_k g^{km} \p_m \phi \r ) \\
& = P^{ab} \p_b P^{ac}  \p_c \phi = \l [ \p^2 - \p_r^2 -(d-1) \p_r \r ] \phi. \nn
\end{align}
This formula allows to find eigenvalues of the scalar Laplacian. Namely, restricting to 
\begin{equation}
\phi =Y_{\ell,m},
\end{equation}
which is given a harmonic homogeneous polynomial (in one-to-one correspondence with symmetric traceless tensors, which are the irreps of the $SO(d)$) of degree $\ell$, we get
\begin{align}
\nabla^2 Y_{\ell,m} = - \l [ \ell(\ell-1) + (d-1) \ell \r ] Y_{\ell,m} = - \ell( \ell+d-2 )Y_{\ell,m}.
\end{align}

Similarly, we can get
\begin{align}
\nabla^i \nabla_i v_j & = g^{ik}\nabla_k \l ( e^a_j e^c_i P^{ab} \p_c v^b \r ) = 
g^{ik}\nabla_k e^a_j e^c_i P^{ab} \p_c v^b + g^{ik} e^a_j \nabla_k e^c_i P^{ab} \p_c v^b+g^{ik} e^a_j e^c_i \p_k \l( P^{ab} \p_c v^b \r ) \nn \\
& = - (d-1) n^c  e^a_j P^{ab} \p_c v^b - e^a_j n^b P^{ac} \p_c v^b + e^a_j P^{ab} P^{cd}\p_c \p_d v^b,
\end{align}
or equivalently,
\begin{equation}
e^a_j \nabla^i \nabla_i v^j = \underbrace{P^{ab} P^{cd}\p_c \p_d v^b}_{(A)} - (d-1) \underbrace{n^c P^{ab} \p_c v^b}_{(B)} - \underbrace{n^b P^{ac} \p_c v^b}_{(C)}.
\end{equation}
Components of the vector $v^a$ are not independent, therefore, it is not this vector that should be considered as a candidate for being given by a harmonic polynomial. Instead, we consider a vector field in the ambient space $V^a$, which can always be expanded as
\begin{equation}
V^a = P^{ab} V^b + n^a n^b V^b = v^a + n^a \s.
\end{equation}
As a result, we have for $(A)$
\begin{align}
(A) & = P^{ab} \p^2 \l ( V^b - n^b \s \r ) - P^{ab} n^c \p_c n^d \p_d v^b + P^{ab} n^c (\p_c n^d) \p_d v^b \\
& = P^{ab} \p^2 V^b - P^{ab} \p_c \l ( P^{bc} \s + n^b \p_c \s \r ) - P^{ab} \p_r^2 v^b 
= P^{ab} \p^2 V^b - 2 P^{ab} \p_b \s - P^{ab} \p_r^2 v^b \nn
\end{align}
for $(B)$
\begin{align}
(B) & = P^{ab} \p_r v^b,
\end{align}
and for $(C)$
\begin{align}
(C) & = P^{ac} \p_c (n^b v^b) - P^{ac} v^b \p_c n^b = - P ^{ab} v^b = -v^a.
\end{align}
Collecting all terms together, we obtain
\begin{equation}
e^a_j \nabla^i \nabla_i v^j = P^{ab} \p^2 V^b - 2 P^{ab} \p_b \s - P^{ab} \p_r^2 v^b - (d-1) P^{ab} \p_r v^b + v^a.
\end{equation}
Similarly, using \eqref{eq:LaplacianSigma}
\begin{align}
\nabla^2 \s & = \p_c  (v^c + n^a \p_c V^a) - \p_r^2 \s -(d-1) \p_r \s 
=  \p_c v^c + P^{ac} \p_c V^a +n^a \p^2 V^a - \p_r^2 \s -(d-1) \p_r \s \nn \\
& = \nabla_i v^i + P^{ac} \p_c \l ( v^a  + n^a \s \r ) + n^ a \p^2 V^a - \p_r^2 \s -(d-1) \p_r \s \nn \\
& = 2 \nabla_i v^i +(d-1) \s + n^ a \p^2 V^a - \p_r^2 \s -(d-1) \p_r \s
\end{align}
Choosing the vector $V^A$ as a homogeneous harmonic polynomial of degree $\ell$,\footnote{It is worth noting that both $v^a$ and $\s$ are also homogeneous polynomials of degree $\ell$.} we obtain
\begin{align}
e^a_j \nabla^i \nabla_i v^j & = - 2 g^{ij} e^a_i \nabla_j \s - \ell(\ell-1) v^a - \ell (d-1) v^a + v^a \nn \\
& = - 2 g^{ij} e^a_i \nabla_j \s - \l [ \ell(\ell+d-2) -1 \r ] v^a,
\end{align}
equivalently
\begin{align}
\label{eq:EigenModesV}
\nabla^i \nabla_i v_j & = - 2 \nabla_j \s - \l [ \ell(\ell+d-2) -1 \r ] v_j,
\end{align}
and
\begin{align}
\label{eq:EigenModesSigma}
\nabla^2 \s & = 2 \nabla_i v^i - \l [ \ell(\ell+d-2) - (d-1) \r ] \s.
\end{align}

To find the eigenvalues of the vector Laplacian, we consider two case. The first, the longitudinal mode, when
\begin{equation}
v_i = \nabla_i \phi.
\end{equation}
In this case, the two equations become
\begin{align}
& \nabla^2 \phi +2 \s + \l [ \ell(\ell+d-2) + d - 3 \r ] \phi = 0, \nn \\
& \nabla^2 \s - 2 \nabla^2 \phi + \l [ \ell(\ell+d-2) - (d-1) \r ] \s = 0.
\end{align}
Looking for a solution to this system in the form
\begin{equation}
\l (
\begin{array}{c}
\phi \\
\s
\end{array}
\r )=
Y_{s,m} \l(
\begin{array}{c}
\tilde \phi \\
\tilde \s
\end{array}
\r),
\end{equation}
with constants $\tilde \phi$ and $\tilde \s$, the condition for having a non-trivial solution is given by
\begin{equation}
\m_s^{(+)} = (\ell+1) (\ell+d-1), ~~ \m_s^{(-)} = (\ell-1) (\ell+d-3).
\end{equation}
We see that these two solutions correspond to
\begin{equation}
s_\pm = \ell \pm 1.
\end{equation}
Therefore, the solution corresponding to spin $\ell$ is given by
\begin{equation}
\label{eq:dV0}
v_i = \nabla_i Y_{\ell m},
\end{equation}
with eigenvalues and degeneracies given by
\begin{equation}
\mu^S_{d,\ell} = \ell(\ell+d-2), ~~ N^S_{d,\ell}= \f{2\ell+d-2}{\ell}
\l (\begin{array}{c}
\ell+d-3 \\
d-2
\end{array}\r)
\end{equation}

The second case corresponds a divergenceless vector
\begin{equation}
\nabla_i u^i = 0.
\end{equation}
It follows immediately from \eqref{eq:EigenModesV} that $\nabla^2 \s=0$, therefore, using \eqref{eq:EigenModesSigma} we conclude that $\s = 0$. As a result, these modes satisfy the following equation
\begin{align}
\nabla^i \nabla_i v_j & = - \l [ \ell(\ell+d-2) -1 \r ] v_j,
\end{align}
meaning that the vector eigenvalues are given by
\begin{equation}
\label{eq:VectorEigenvalues}
\m^V_{d,\ell} = \ell(\ell+d-2) -1.
\end{equation}
To find the degeneracy we proceed as follows. A generic homogeneous harmonic vector is given by
\begin{equation}
V^a = T_{a a_1 \dots a_\ell} x^{a_1}\dots x^{a_\ell},
\end{equation}
the constraints \eqref{eq:dV0} and $\s=0$ in the embedding space translate into
\begin{equation}
V^a n^a = 0, ~~ \p_a V^a = 0.
\end{equation}
The solution to these equation is a tensor $T_{a(a_1\dots a_\ell)}$ with a hook symmetry. Counting the number of these tensors we arrive to the following formula for the degeneracy of the vector harmonics with eigenvalues
\begin{equation}
N^V_{d,\ell} = \f{\ell(2\ell+d-2)}{\ell+d-3}
\l (\begin{array}{c}
\ell+d-2 \\
d-3
\end{array}\r)
\end{equation}

\section{Geometric relations}

All commutators can be obtained using the following formula
\begin{align}
[\nabla_i, \nabla_j] A^k & = A^m R^{k}_{mij},
\end{align}
and that for the sphere $\mathbb S^{d-1}$ the curvature is given by
\begin{equation}
R_{ijk\ell} = g_{ik} g_{j\ell} - g_{i\ell} g_{jk}, \quad R_{j\ell} = (d-2) g_{j \ell}, \quad R = (d-1)(d-2).
\end{equation}
In particular we get for three derivatives
\begin{align}
\nabla_i \nabla^2 \phi & = \nabla_i \nabla^2 \phi + (d-2) \nabla_i \phi,
\end{align}
for four derivatives
\begin{align}
\nabla_i \nabla_j \nabla^i \nabla^j \phi & = \nabla_k \nabla^2 \nabla^k \phi = (\nabla^2)^2 \phi +(d-2) \nabla^2\phi, \\
\nabla^2 \nabla_i \nabla_j \phi & = \nabla_i \nabla_j \nabla^2 \phi + 2(d-1) \nabla_i \nabla_j \phi -2 g_{ij} \nabla^2 \phi,
\end{align}
and for six derivatives
\begin{align}
\nabla_k (\nabla^2)^2 \nabla^k \phi & = (\nabla^2)^3 \phi + 2 (d-2) (\nabla^2)^2 \phi + (d-2)^2 \nabla^2 \phi, \\
\nabla_i \nabla_j \nabla^2 \nabla^i \nabla^j \phi & 
= (\nabla^2)^3 \phi + 3 (d-2) \nabla^2 \phi + 2 (d-1) (d-2) \nabla^2 \phi.
\end{align}

\newpage
\appendix

\bibliographystyle{JHEP}
\bibliography{LargeChargeBootstrap}{}

\end{document}